\documentclass[pra, reprint, 10pt, english,british]{revtex4-1}
\usepackage[T1]{fontenc}
\usepackage[latin9]{inputenc}
\usepackage[letterpaper]{geometry}
\usepackage{color}
\usepackage{amsmath}
\usepackage{amssymb}
\usepackage{graphicx}
\usepackage{esint}

\makeatletter

\newcommand{\noun}[1]{\textsc{#1}}
\providecommand{\tabularnewline}{\\}
\makeatother

\usepackage{babel}
\begin{document}

\title{Complete Photoionization Experiments via Ultrafast Coherent Control
with Polarization Multiplexing II: \\
Numerics \& Analysis Methodologies}

\author{P. Hockett}

\email{paul.hockett@nrc.ca}

\affiliation{National Research Council of Canada, 100 Sussex Drive, Ottawa,K1M
1R6, Canada}

\author{M. Wollenhaupt}

\affiliation{Institut für Physik, Carl von Ossietzky Universität Oldenburg, Carl-von-Ossietzky-Straße
9-11, 26129 Oldenburg, Germany}

\author{C. Lux }

\author{T. Baumert}

\affiliation{Institut für Physik, Universität Kassel, Heinrich-Plett-Str. 40,
34132 Kassel, Germany}
\begin{abstract}
The feasibility of complete photoionization experiments, in which
the full set of photoionization matrix elements are determined, using
multiphoton ionization schemes with polarization-shaped pulses has
recently been demonstrated {[}Hockett et. al., Phys. Rev. Lett. 112,
223001 (2014){]}. Here we extend on our previous work to discuss further
details of the numerics and analysis methodology utilised, and compare
the results directly to new tomographic photoelectron measurements,
which provide a more sensitive test of the validity of the results.
In so doing we discuss in detail the physics of the photoionziation
process, and suggest various avenues and prospects for this coherent
multiplexing methodology.
\end{abstract}
\maketitle

\section{Introduction}

The aim of ``complete'' photoionization studies is the determination
of the amplitudes and phases of the ionization matrix elements, which
constitute a fundamental description of an ionization event \cite{Reid2003,Cherepkov2005}.
The matrix elements define the coupling of the initial state to the
final compound state, comprised of an ion and free electron. In the
dipole limit, this matrix element can be very generally defined as
$\langle\psi^{e};\,\Psi^{+}|\hat{\boldsymbol{\mu}}.\mathbf{E}|\Psi\rangle$.
Here $\Psi$ is the initial wavefunction of the system, $\Psi^{+}$
the photoion, $\psi^{e}$ the photoelectron, $\hat{\boldsymbol{\mu}}$
the dipole operator and $\mathbf{E}$ the electric field. By expressing
the continuum wavefunction $\psi^{e}$ as a set of \emph{partial-waves},
corresponding to different continuum angular momentum states, the
ionization matrix element can be decomposed into various geometric
and radial components, and the set of amplitudes and phases of these
components constitutes a complete description of the ionization event.
In order to determine these matrix elements from experimental data,
an observable sensitive to the relative phases of the partial-waves
is required, and such an interferometric observable is found in the
photoelectron angular distributions (PADs), which are angular interference
patterns dependent on the composition of $\psi^{e}$.

A range of experiments have been performed in order to provide such
complete descriptions of photoionization for a number of atomic and
molecular systems. The key concern in such experiments is the level
of detail required in order to undertake the relatively complex analysis
procedure.  Typically the angular (or geometric) part of the matrix
elements can be calculated analytically \cite{Dill1976}, leaving
only the energy-dependent radial (or dynamical) components to be determined
from the experimental data. The determination of these components
involves fitting experimental data with the specific ionization formalism
for the ionization event under study. Since, in general, there may
be many partial-waves and the composition of $\psi^{e}$ is not usually
known \emph{a priori}, a large experimental dataset is required for
this procedure. In order to obtain a sufficient dataset, experimental
data is obtained for a range of geometric parameters, for example
by varying the polarization state and polarization geometry \cite{Lambropoulos1973,berry1976,Duong1978,Hansen1980,Chien1983}
or, for molecules, the rotational state or axis distribution \cite{Reid1991,Reid1992,Suzuki2006,Hockett2009,Suzuki2012},
or via molecular frame measurements \cite{Gessner2002,Lebech2003,Yagishita2005}.
Since the dynamical parameters are invariant to these geometric changes,
a dataset of sufficient information content to determine these parameters
may be obtained in this way.

Recently, we demonstrated a new type of measurement and analysis methodology
for complete experiments \cite{Hockett2014}. This method can be considered
as \emph{time-domain polarization-multiplexing}. In this case, a multiphoton
ionization scheme with a moderately intense, ultrafast laser pulse
was employed to ionize potassium atoms. The resulting light-matter
interaction can be understood as an intra-pulse two-step processes,
in which electronic population transfer is driven by the laser field
(i.e. Rabi oscillations), and the excited state population created
can subsequently be ionized via 2-photon absorption. In this case,
the population dynamics and the ionization dynamics are dependent
on the properties of the laser pulse, as well as the physical properties
of the system which ultimately determine the matrix elements. In this
scheme, changing the polarization of the pulse corresponds to changing
the geometric parameters of the ionization, as described above. In
the simplest case a single or pure polarization state is employed,
and the geometric parameters are time-invariant. More generally, via
the use of a polarization-shaped pulse, the geometric parameters can
be changed in a time-dependent manner. Since the dynamics and ionization
all occur within a single laser pulse, the process is fully-coherent,
and the final, time-integrated, photoelectron measurement can be considered
as a time-domain multiplexed measurement of the set of (instantaneous)
polarization states explored by the shaped-pulse.

Here we discuss further details of the work presented in ref. \cite{Hockett2014},
with a focus on extending the details of the theory presented therein,
in particular the numerical details of the fitting procedure and a
discussion of the benefits and limitations of this approach. We further
present detailed comparison of our results with new maximum information
photoelectron measurements, utilizing a tomographic procedure for
the measurement of 3D photoelectron distributions and detailed analysis,
allowing for a quantitative comparison of the predicted PADs and experimental
PADs as a function of polarization geometry (further details of the
maximum information measurements can be found in ref. \cite{Hockett2015b}).
Finally, the possibilities of extending this treatment to different
classes of ionization is explored, with a particular emphasis on molecular
ionization problems.

\section{Intrapulse dynamics \& multiphoton ionization with polarization-shaped
pulses\label{sec:Intrapulse-dynamics-theory}}

Here we detail the various steps involved in the treatment of the
3-photon scheme detailed above. For completeness we include all aspects
of our treatment.

\begin{figure}
\includegraphics{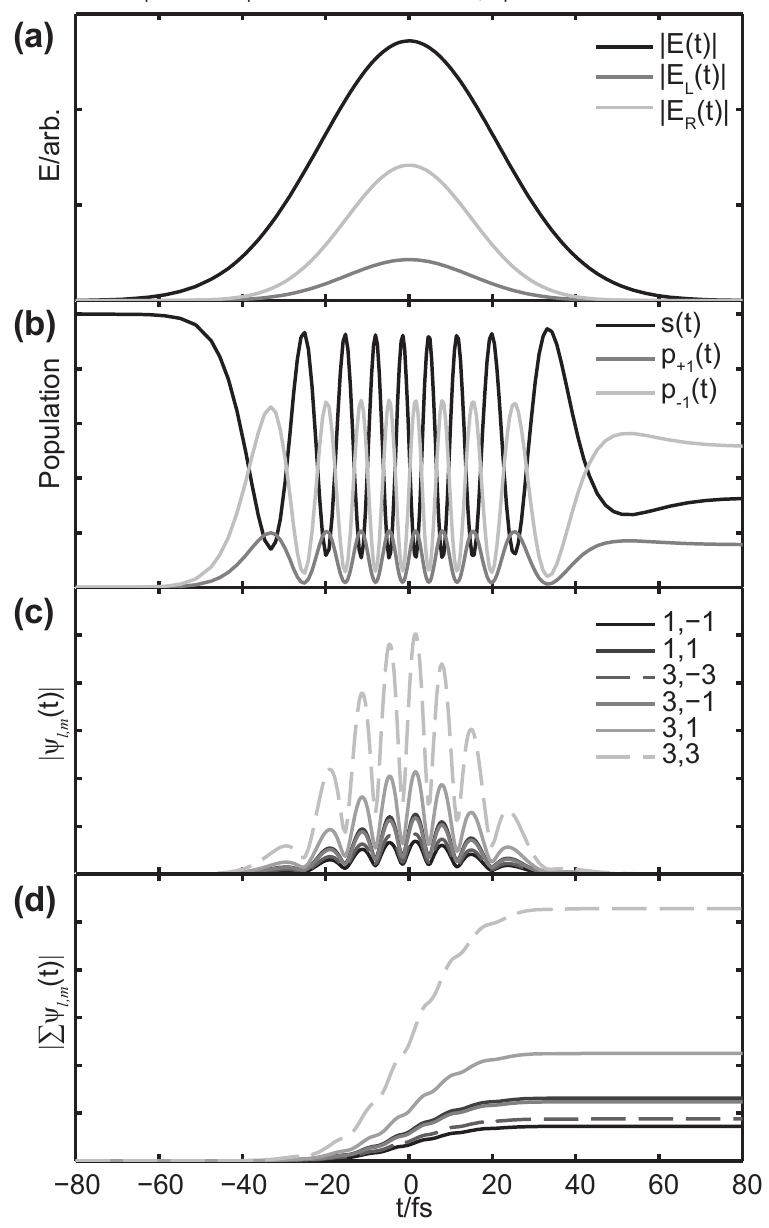}

\caption{%
Time-dependent dynamics for an elliptically polarized laser pulse.
(a) Laser field envelope for an elliptically polarized pulse, defined
by $\phi_{y}=0.5$~rad (eqn. \ref{eq:E-spherical}). (b) Bound state
populations (eqn. \ref{eq:TDSE}). (c) Instantaneous continuum populations
$d_{l_{f},m_{f}}(k,t)$ (eqn. \ref{eq:dlmt}) and (d) cumulative continuum
population (eqn. \ref{eq:dtInt}).\label{fig:Time-dependent-dynamics-elliptical}%
}
\end{figure}

\begin{figure}
\includegraphics{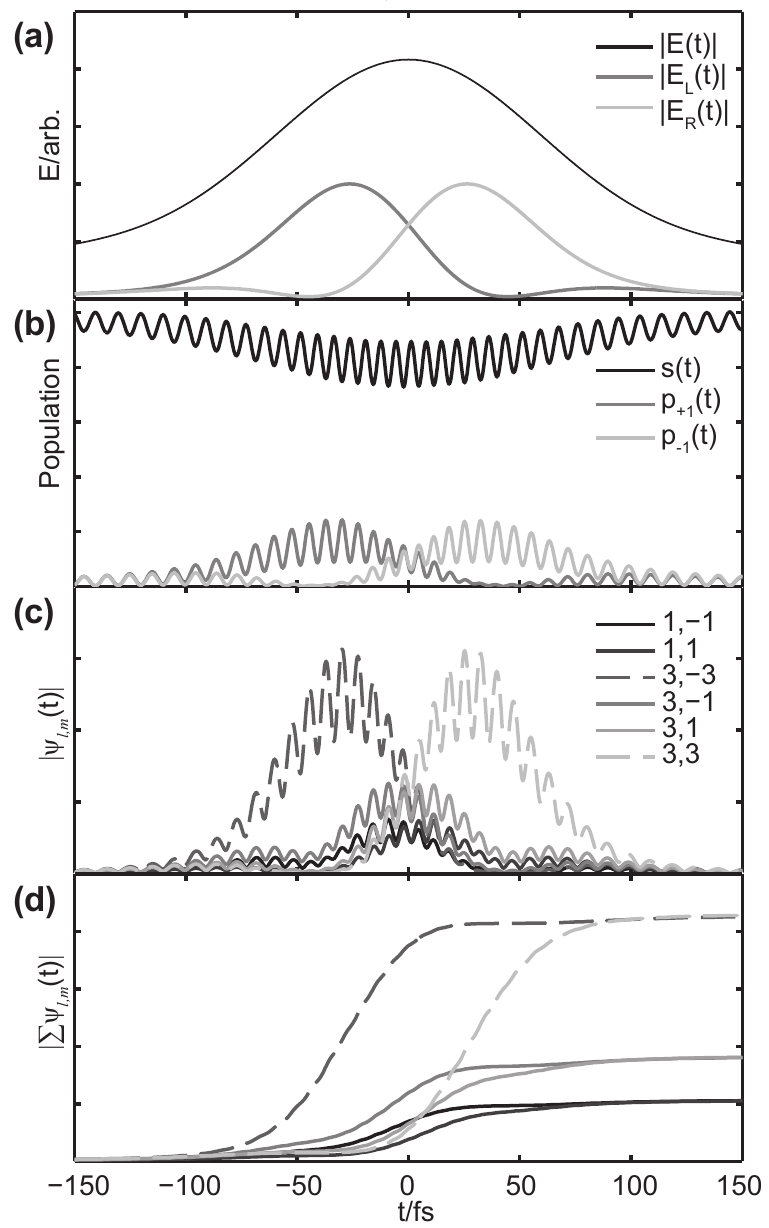}

\caption{
Time-dependent dynamics for a polarization-shaped laser pulse. (a)
Laser field envelope for an elliptically polarized pulse, defined
by $\phi_{y}=-\pi$~rad for the blue half of the pulse (eqn. \ref{eq:E-spherical}).
(b) Bound state populations (eqn. \ref{eq:TDSE}). (c) Instantaneous
continuum populations $d_{l_{f},m_{f}}(k,t)$ (eqn. \ref{eq:dlmt})
and (d) cumulative continuum population (eqn. \ref{eq:dtInt}).\label{fig:Time-dependent-dynamics-shaped}%
}

\end{figure}

\subsection{Electric field}

The electric field as a function of time is described as:

\begin{equation}
E(t)=E_{0}e^{-(t/\tau)^{2}}e^{i\omega t}
\end{equation}
where $E_{0}$ is the field strength, the pulse envelope is Gaussian
with temporal width parameter $\tau$ and $\omega$ is the carrier
(angular) frequency. Using the notation of ref. \cite{dielsandrudolph}
the spectral content of the pulse is given by:

\begin{equation}
\tilde{E}(\Omega)=\mathcal{F}\{E(t)\}
\end{equation}
where $\mathcal{F}$ represents a Fourier Transform. 

Polarization shaped pulses are described as in ref. \cite{Wollenhaupt2009a},
by assuming initially identical and in-phase $(x,y)$ field components,
then applying a spectral phase shift. Hence, a field described by
two Cartesian components with independent spectral phases (but identical
spectral content) can be defined as:

\begin{equation}
\left(\begin{array}{c}
\tilde{E_{x}}(\Omega)\\
\tilde{E_{y}}(\Omega)
\end{array}\right)=\tilde{E}(\Omega)\left(\begin{array}{c}
e^{i\phi_{x}(\Omega)}\\
e^{i\phi_{y}(\Omega)}
\end{array}\right)
\end{equation}
resulting in the time-domain components:

\begin{equation}
\left(\begin{array}{c}
E_{x}(t)\\
E_{y}(t)
\end{array}\right)=\mathcal{F}^{-1}\left\{ \tilde{E}(\Omega)\left(\begin{array}{c}
e^{i\phi_{x}(\Omega)}\\
e^{i\phi_{y}(\Omega)}
\end{array}\right)\right\} 
\end{equation}

The field can also be expressed in terms of a spherical basis, i.e.
left and right circularly polarized components:

\begin{equation}
\left(\begin{array}{c}
E_{L}(t)\\
E_{R}(t)
\end{array}\right)=\frac{1}{\sqrt{2}}\left(\begin{array}{c}
E_{x}(t)-iE_{y}(t)\\
E_{x}(t)+iE_{y}(t)
\end{array}\right)\label{eq:E-spherical}
\end{equation}
This final form was used in the calculations herein, since it physically
describes the instantaneous pulse angular momentum, in terms of the
projection of the photon momentum onto the propagation axis, where
$L$ equates to $m=+1$ and $R$ to $m=-1$ states. This form can
therefore be directly interpreted in terms of the allowed $\Delta m$
of both bound-bound and bound-free transitions - this is discussed
further below. Note that this form implies that the light propagates
along the $z$-axis, and the lab. frame angular momentum $m$ is defined
relative to this propagation axis.

\subsection{Non-perturbative laser-atom interaction}

The strong laser field drives Rabi oscillations in the atom, coupling
electronic states $|n,l,m\rangle$. In the case of potassium atoms,
as detailed in ref. \cite{Wollenhaupt2009a}, the initial population
is in the 4$s$ state and the laser frequency is near resonant with
the $|4,0,0\rangle\rightarrow|4,1,m\rangle$ transition, hence single
photon absorption populates the 4$p$ manifold, while a strong laser
field will drive Rabi cycling between the $4s$ and $4p$ states.
The allowed values of $m$ depend on the polarization state of the
light.

The population dynamics during the laser pulse, described in the spherical
basis of eqn. \ref{eq:E-spherical}, are given by the time-dependent
Schrödinger equation:

\begin{equation}
\frac{d}{dt}\left(\begin{array}{c}
s(t)\\
p_{+1}(t)\\
p_{-1}(t)
\end{array}\right)=i\left(\begin{array}{ccc}
0 & \frac{1}{2}\Omega_{L}^{*}(t) & \frac{1}{2}\Omega_{R}^{*}(t)\\
\frac{1}{2}\Omega_{L}(t) & \delta_{+1} & 0\\
\frac{1}{2}\Omega_{R}(t) & 0 & \delta_{-1}
\end{array}\right)\left(\begin{array}{c}
s(t)\\
p_{+1}(t)\\
p_{-1}(t)
\end{array}\right)\label{eq:TDSE}
\end{equation}
where $s(t)$, $p_{+1}(t)$ and $p_{-1}(t)$ are the state vector
components for the $|4,0,0\rangle$ and $|4,1,\pm1\rangle$ states,
$\Omega_{L/R}(t)=\mu_{L/R}E_{L/R}(t)$ where $\mu_{L/R}$ are the
transition amplitudes, and $\delta_{\pm1}$ represent the detuning
of the laser from the resonant frequency of the transition. Here it
is clear that the $L$ and $R$ components of the electric field drive
transitions with $\Delta m=+1$ and $\Delta m=-1$ respectively; this
is simply the consequence of the conservation of angular momentum
since the light carries $l=1$ unit of angular momentum, with lab.
frame projection $m=1$ for $E_{L}$ and $m=-1$ for $E_{R}$. In
this sense the (instantaneous) helicity of the electric field is directly
imprinted on the atomic ensemble.

Here $\hbar$ and $E_{0}$ are both set to unity for simplicity; $\mu_{L/R}$
is also set to unity, i.e. equal probability of transitions to both
$|4,1,m\rangle$ states, and $\delta_{\pm1}=0.05$~rad/fs. For determination
of PADs these simplifications are acceptable as only the \emph{relative}
population of $m=\pm1$ states will affect the angular distribution,
and these populations are dependent only on the driving laser field
polarization.

\subsection{Perturbative two-photon ionization \& PADs}

In the perturbative regime, the dipole transition amplitude for a
transition from a bound state $|n_{i},l_{i},m_{i}\rangle$ to a continuum
state $|\mathbf{k};\, l_{f},m_{f}\rangle$ is given by the dipole
matrix elements:

\begin{eqnarray}
d_{i\rightarrow f}(k,t) & = & \langle\mathbf{k};\, l_{f}m_{f}|\hat{\boldsymbol{\mu}}_{if}.E(t)|n_{i}l_{i}m_{i}\rangle\\
 & \propto & R_{l_{i}l_{f}}^{n}(k)E_{q}(t)\langle l_{f}m_{f},1q|l_{i}m_{i}\rangle\label{eq:d_if-kt}
\end{eqnarray}
where $\boldsymbol{\hat{\mu}}_{if}$ is the dipole operator; $R_{l_{i}l_{f}}^{n}(k)$
is the radial part of the matrix element, which is dependent on the
magnitude of the photoelectron wavevector $\mathbf{k}$, the principal
quantum number of the initial state $n$ and the electronic orbital
angular momentum $l$, but assumed to be independent of $m_{i}$ and
$m_{f}$; $\langle l_{f}m_{f},1q|l_{i}m_{i}\rangle$ is a Clebsch-Gordan
coefficient which describes the angular momentum coupling for single
photon absorption, with $q=\pm1$ for the $L$ and $R$ components
of the laser field respectively. This treatment corresponds to a single
active electron picture, in which the final state is a pure continuum
state, i.e. the photoion is neglected and there is no angular momentum
transfer to core. Spin is also neglected. This treatment is sufficient
for the potassium atom case discussed herein; extension to more complex
coupling schemes is discussed in sect. \ref{sub:Assumptions-&-extensions}.

Under these assumptions, the angular part of both bound-bound and
bound-free transitions are described by matrix elements of the same
form. Using these dipole matrix elements, two-photon ionization to
a single final state $|l_{f},m_{f}\rangle$, from an initial state
$|n_{i},l_{i},m_{i}\rangle$, via a virtual one-photon state $|n_{v},l_{v},m_{v}\rangle$,
can then be written as:

\begin{widetext}

\begin{eqnarray}
d_{l_{f}m_{f}}(k,t) & = & d_{i\rightarrow v}(k,t)d_{v\rightarrow f}(k,t)\\
 & = & \sum_{\begin{array}{c}
n_{i},l_{i},m_{i}\\
n_{v},l_{v},m_{v}\\
q,q'
\end{array}}R_{l_{v}l_{f}}^{n_{v}}(k)E_{q'}(t)\langle l_{f}m_{f},1q'|l_{v}m_{v}\rangle R_{l_{i}l_{v}}^{n_{i}}(k)E_{q}(t)\langle l_{v}m_{v},1q|l_{i}m_{i}\rangle\chi_{n_{i},l_{i},m_{i}}(t)
\end{eqnarray}
\end{widetext}This form shows the general case, with summation over
all initial states $|n_{i},l_{i},m_{i}\rangle$ weighted by their
populations $\chi_{n_{i},l_{i},m_{i}}(t)$. Although the bound-free
matrix element is labelled with quantum number $n_{v}$, in practice
this is unassigned and will correspond to a quasi-continuum of virtual
states within the laser bandwidth, so is dropped in the following.%
\footnote{For completeness we note that in the presence of resonances at the
1-photon level, the bound-bound transitions would look identical within
a single active electron model, apart from taking on specific, well-defined
values of $n$. In the case where several resonant states, e.g. high-lying
Rydbergs, were within the laser bandwidth the dependence of the magnitudes
and phases on $n$ would be significant. Pertinent examples of this
type of effect in a multi-photon ionization scheme can be found in
refs. \cite{Krug2009,Wilkinson2014}.%
} In this treatment all energy dependence is contained in the $R(k)$
radial integrals. For the potassium case considered here, a slightly
simplified form can be written since only the 4$p$ levels contribute
to the ionization, hence $n_{i}=4$, and the time-dependent populations
are given by $p_{m_{i}}(t)$ as defined in eqn. \ref{eq:TDSE}:

\begin{widetext}

\begin{eqnarray}
d_{l_{f}m_{f}}(k,t) & = & \sum_{\begin{array}{c}
l_{i},m_{i}\\
l_{v},m_{v}\\
q,q'
\end{array}}R_{l_{v}l_{f}}(k)E_{q'}(t)\langle l_{f}m_{f},1q'|l_{v}m_{v}\rangle R_{l_{i}l_{v}}^{(4)}(k)E_{q}(t)\langle l_{v}m_{v},1q|l_{i}m_{i}\rangle p_{m_{i}}(t)\label{eq:dlmt}
\end{eqnarray}

Integrating over $t$ yields:

\begin{equation}
d_{l_{f}m_{f}}(k)=\int d_{i\rightarrow v}(k,t)d_{v\rightarrow f}(k,t)dt=\int dt\sum_{\begin{array}{c}
l_{i},m_{i};l_{v},m_{v}\\
q,q'
\end{array}}R_{l_{v}l_{f}}(k)\langle l_{f}m_{f},1q'|l_{v}m_{v}\rangle R_{l_{i}l_{v}}^{(4)}(k)\langle l_{v}m_{v},1q|l_{i}m_{i}\rangle E_{q'}(t)E_{q}(t)p_{m_{i}}(t)\label{eq:dtInt}
\end{equation}

The observed photoelectron yield as a function of angle, the PAD,
for a small energy range $dk$ over which we assume the $R(k)$ are
constant, is then given by the coherent square over all final (photoelectron)
states:

\begin{eqnarray}
I(\theta,\phi;\, k) & =\int dk & \sum_{\begin{array}{c}
l_{f},m_{f}\\
l_{f}^{'},m_{f}^{'}
\end{array}}d_{l_{f}m_{f}}(k)Y_{l_{f}m_{f}}(\theta,\phi)d_{l_{f}^{'}m_{f}^{'}}^{*}(k)Y_{l_{f}^{'}m_{f}^{'}}^{*}(\theta,\phi)\label{eq:Itp}
\end{eqnarray}

\end{widetext}

This treatment is very similar to that given in ref. \cite{Wollenhaupt2009a},
with the Clebsch-Gordan coefficients equivalent to the $\alpha_{l,m;l',m'}$
parameters and the $d_{l_{f}m_{f}}$ similar to the $c_{l,m}$. The
main difference is that all $|l_{f},m_{f}\rangle$ are accounted for,
hence the explicit inclusion of the radial elements $R_{ll}(k)$.
The radial matrix elements defined here are assumed to be complex,
and include both the scattering phase $e^{-i\eta_{l}}$ and the geometric
phase factor $i^{l}$ which usually appear in the definition of the
photoelectron wavefunction \cite{Park1996}. The amplitudes and phases
of these parameters constitute the unknowns which are sought in ``complete''
photoionization studies and, physically, define the scattering of
the outgoing photoelectron from the nascent ion core.

The PAD can also be described by a generic expansion in spherical
harmonics with expansion coefficients $\beta_{L,M}$, termed anisotropy
parameters, where:

\begin{equation}
I(\theta,\phi;\, k)=\sum_{L,M}\beta_{L,M}(k)Y_{L,M}(\theta,\,\phi)\label{eq:IBlm}
\end{equation}
In general the $\beta_{L,M}(k)$ provide a compact way to express
the PADs, and allowed values are constrained by symmetry \cite{Yang1948,Reid2003}.
This expansion can be considered as indicating the information content
of a given distribution, and the resultant multipole moments $L,\, M$
are related to the partial wave expansion of eqn. \ref{eq:IBlm} by
\cite{Reid1991}:

\begin{widetext}

\begin{equation}
\beta_{L,M}=\underset{\begin{array}{c}
l_{f},m_{f}\\
l_{f}^{'},m_{f}^{'}
\end{array}}{\sum}\sqrt{\frac{(2l_{f}+1)(2l'_{f}+1)(2L+1)}{4\pi}}\left(\begin{array}{ccc}
l_{f} & l'_{f} & L\\
0 & 0 & 0
\end{array}\right)\left(\begin{array}{ccc}
l_{f} & l'_{f} & L\\
m_{f} & -m'_{f} & M
\end{array}\right)d_{l_{f}m_{f}}(k)d_{l_{f}^{'}m_{f}^{'}}^{*}(k)
\end{equation}

\end{widetext}

Further exploration of the information content of PADs for the case
of tomographic 3D photoelectron measurements can be found in ref.
\cite{Hockett2015b}.

\subsection{Pure and shaped laser pulse dynamics}

In order to illustrate the theory detailed above, figures \ref{fig:Time-dependent-dynamics-elliptical}
and \ref{fig:Time-dependent-dynamics-shaped} give the details of
two example calculations, for an elliptically polarized pulse and
a fully polarization-shaped pulse respectively. In both cases the
panels illustrate, from top to bottom, the envelope of the laser field
and $L$, $R$ components, as defined by eqn. \ref{eq:E-spherical};
the population dynamics driven by the laser field, in terms of the
state vector components $s(t)$, $p_{+1}(t)$ and $p_{-1}(t)$ defined
in eqn. \ref{eq:TDSE}; the instantaneous continuum populations, as
defined by eqn. \ref{eq:dlmt} and making use of the previously determined
photoionization matrix elements $R_{ll}$ (see ref. \cite{Hockett2014}
and sect. \ref{sec:Photoelectron-image-generation}); the cumulative
continuum populations, as defined by eqn. \ref{eq:dlmt}.

Both examples provide insight into the dynamics of the ionization
process, and it is clear how the $L$ and $R$ components of the laser
field drive both the bound-state population dynamics, and the instantaneous
continuum contributions. Since, in this model, the two steps are decoupled,
and the ionization is assumed to be perturbative, there is no depletion
in the bound state populations. The ionization step does, however,
follow the bound state dynamics since the instantaneous population
defines which continuum states can be accessed, and their relative
weighting. Thus, the instantaneous continuum dynamics follow the bound-state
dynamics. Furthermore, since there are no continuum electron dynamics
in this model (i.e. no laser-continuum coupling, or electron-ion recombination),
the final continuum is simply the sum over the instantaneous continuum
contributions (eqn. \ref{eq:dtInt}) and builds up coherently over
the pulse envelope. The resultant PAD, eqn. \ref{eq:Itp}, thus depends
both on the final continuum populations, as well as the accumulated
phase for each $|l_{f},m_{f}\rangle$ state.

In the case of a ``pure'' polarization state (fig. \ref{fig:Time-dependent-dynamics-elliptical}),
in this example an elliptically polarized light field defined by $\phi_{y}=0.5$~rad.,
there is essentially no dynamic contribution to the final result since
the \emph{relative} continuum contribution is time-independent. In
the language used previously, the geometric contribution to the ionization
is time-invariant. However, in the case of a polarization-shaped pulse
(fig. \ref{fig:Time-dependent-dynamics-shaped}), where the relative
$L$ and $R$ components do vary significantly over the pulse, the
intra-pulse dynamics play a key role in defining the final continuum
wavefunction. It is this dependence that makes the final PAD particularly
sensitive to the pulse shape, as well as the ionization matrix elements.
While the two cases are formally identical, there is clearly no polarization
multiplexing in the pure case, since the polarization state is time-invariant.
In the polarization-shaped case, the information content is greatly
increased since the final result arises from coherent addition over
all instantaneous polarization states, thus contains additional information
relative to a pure case. (Further examples of polarization-shaped
pulses and resultant PADs can be found in ref. \cite{Hockett2014}.)

\section{Photoelectron image generation \& fitting\label{sec:Photoelectron-image-generation}}

In this section we outline salient details of the numerics used in
applying the above theory to the generation of photoelectron momentum
distributions which can be compared with experimental data. In the
context of complete photoionization experiments, the use of these
momentum distributions to generate 2D photoelectron images and fit
experimental data is described.

\subsection{Photoelectron momentum distributions}

\begin{figure*}
\includegraphics{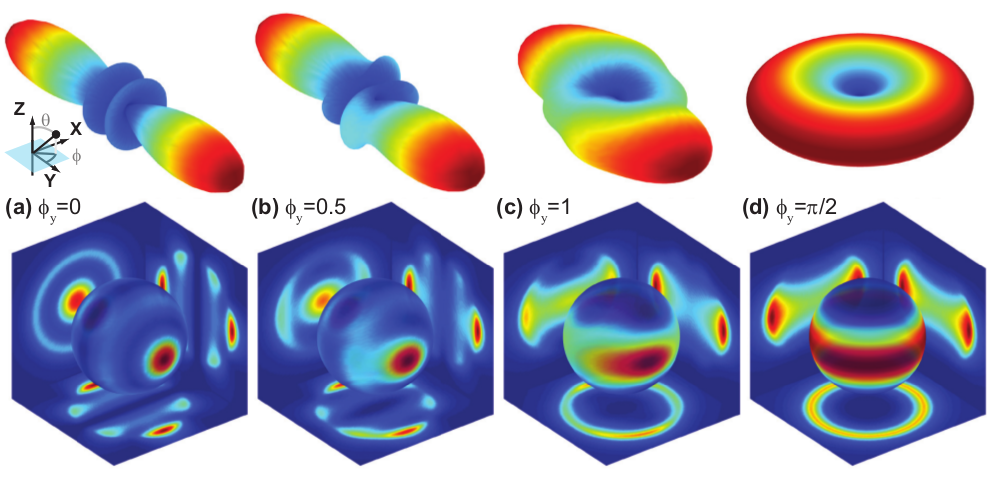}

\caption{\selectlanguage{english}%
Illustration of computed PADs and 2D photoelectron images as a function
of pulse polarization defined by $\phi_{y}$, propagating in the $z$-direction.
The top row shows PADs in polar form, $I(\theta,\,\phi;\, k)$, the
bottom row shows the same distributions projected onto a velocity
iso-sphere (single $k$), and 2D image plane projections $I_{2D}(\theta_{2D},\, k)$
assuming a Gaussian energy spectrum.\label{fig:computed-PADs}%
}

\end{figure*}

The theory detailed above provides a definition of the photoelectron
yield as a function of time, energy and angle, most compactly defined
by the $\beta_{L,M}(k)$ parameters, but ultimately depending on the
underlying laser and target properties. The generation of theoretical,
time-integrated, photoelectron momentum distributions from these parameters
simply involves the population of a 3D grid $(\theta,\phi,k)$ with
the relevant basis set expansion in spherical harmonics as a function
of energy, as defined in eqn. \ref{eq:Itp} (the radial aspect of
this expansion is discussed below).

The volumetric data defined in this way is equivalent to the experimental
data recorded in a 3D imaging experiment, examples of such experiments
are direct 3D imaging via techniques with high temporal and spatial
resolution (for instance refs. \cite{Continetti2001,Reid2012,Hockett2013}
and references therein), or indirect methods based on tomography in
which 3D distributions are reconstructed from a set of 2D projections
\cite{Wollenhaupt2009,Smeenk2009,Hockett2010} (see also ref. \cite{Hockett2015b}).
For comparison with 2D imaging data, further integration along a spatial
dimension is additionally required in order to project the volumetric
data onto a 2D plane. We note that in both imaging experiments and
the numerics applied here, this summation is treated incoherently.
Physically, this corresponds to a loss of photoelectron coherence
before or at the detector, effectively long after the coherent quantum
mechanical scattering event which determines the momentum distribution
(PADs and energy spectrum) \cite{Wollenhaupt2002,Wollenhaupt2013}.
Since the range of the initial scattering event is microscopic, while
photoelectron propagation and detection is macroscopic and often involves
the application of external fields and, ultimately, discrete particle
counting, this is a physically reasonable assumption.

In the results shown in paper I we additionally assumed that the radial
dependence of the ionization matrix elements over the span of the
main spectral feature (\textasciitilde{}200~meV) was negligible,
and that the details of the radial distribution could be simplified
to a Gaussian energy spread with no phase contribution. This allowed
for the momentum data generation and fitting to be simplified, and
the radial distribution given by a Gaussian (defined in energy-space):

\begin{equation}
G(k)=\frac{I_{0}}{\sqrt{2\pi\gamma}}e^{-(E(k)-E(k_{0}))^{2}/2\gamma^{2}}
\end{equation}
where $I_{0}$ is the intensity, $\gamma$ the width, $E(k)$ and
$E(k_{0})$ define the radial coordinate and the peak centre in energy-space.
The final 3D momentum distribution is then defined by: 

\begin{equation}
I(\theta,\phi,k)=G(k)\sum_{L,M}\beta_{L,M}^{k}Y_{L,M}(\theta,\phi)\label{eq:IBlmk}
\end{equation}
Where the $\beta_{L,M}^{k}$ include the superscript to denote that
these parameters are generally dependent on $k$, as in eqn. \ref{eq:IBlm},
but are here taken to be constant over the range of $k$ spanned by
the Gaussian envelope $G(k)$. Finally, it is of note that more generally
the Gaussian assumed here should be replaced by an accurate energy
spectrum, this point is discussed in sect. \ref{sub:Assumptions-&-extensions}.

The 2D images obtained by integration of the volumetric distribution
function are then given as:

\begin{equation}
I_{2D}(\theta_{2D},k)=\intop_{u}I(\theta,\phi,k)\mathrm{d}u\label{eq:I2D}
\end{equation}
where $u$ defines the domain of integration (with integration over
the Cartesian $X$, $Y$ or $Z$ directions for the corresponding
$(y,z)$, $(x,z)$ or $(x,y)$ image planes respectively), and $\theta_{2D}$
is defined in the image plane.

Figure \ref{fig:computed-PADs} illustrates the computed PADs, obtained
using the matrix elements of ref. \cite{Hockett2014}, for four polarization
states of the electric field. The top row shows the PADs in spherical
polar form, as defined by eqn. \ref{eq:IBlm}, while the bottom row
shows the same PADs projected onto spherical surfaces. This is how
the distributions appear in velocity space, as the angle-dependent
photoelectron flux for each $k$. The 2D projections show the same
angular distributions, combined with a Gaussian energy spectrum as
per eqn. \ref{eq:IBlmk}, and projected onto 2D Cartesian planes.
These image planes simulate velocity map imaging data, and illustrate
how the experimental results will depend on both the native details
of the PAD and the details of the projection geometry. In this case,
the laser propagates along the $Z$-axis, and the polarization is
defined in the $(X,Y)$ plane, so experimental images will correspond
to the image planes $(x,z)$ or $(y,z)$ (since images cannot be obtained
in the propagation direction in a standard VMI experiment), and the
precise details will further depend on the rotation of the distribution
about the $Z$-axis. (Further details of 2D and 3D imaging, geometry
considerations and information content, can be found in ref. \cite{Hockett2015b}.)

\subsection{Fitting methodology}

\begin{figure}
\includegraphics[scale=0.8]{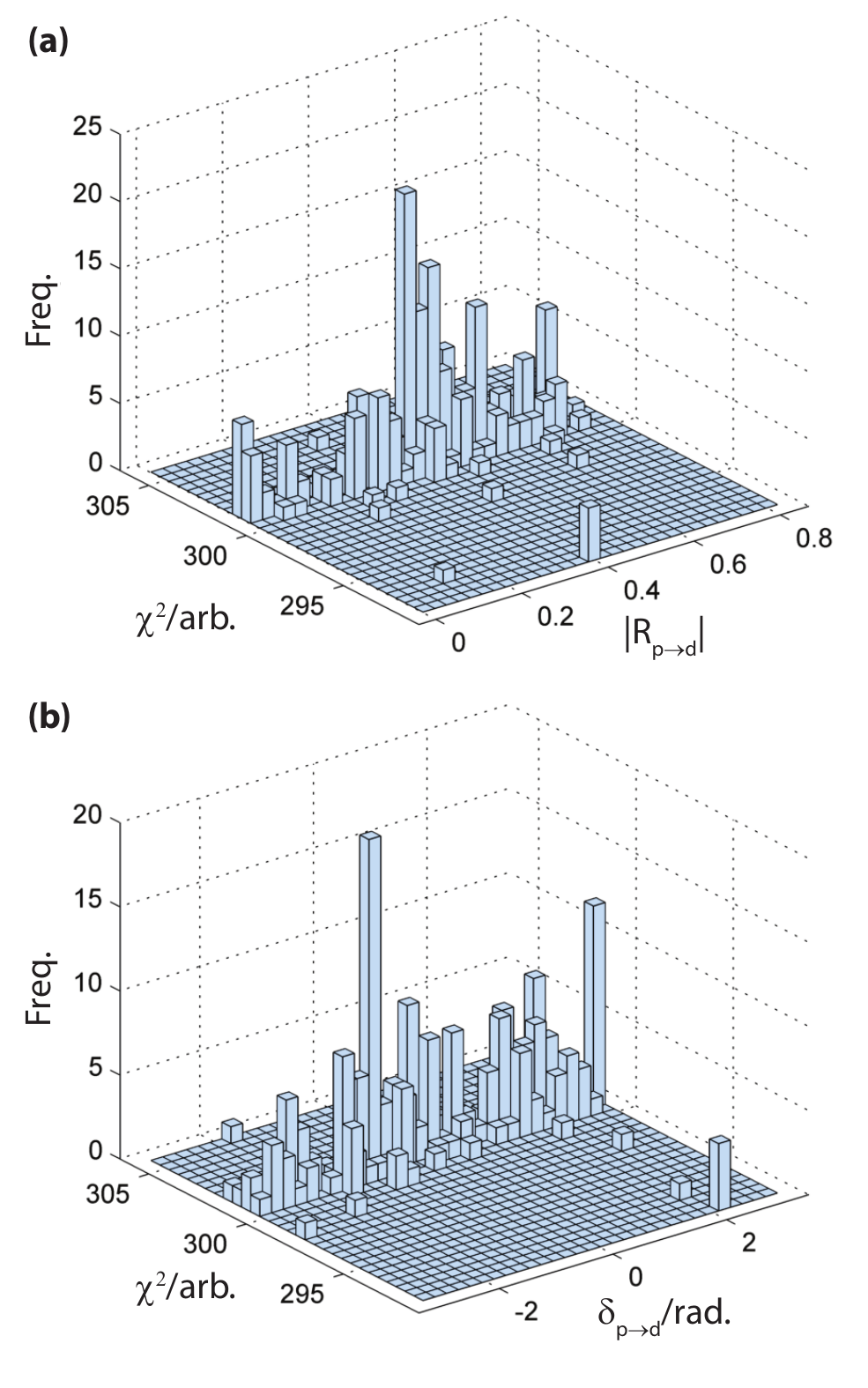}

\caption{%
Example of fit statistics for $R_{p\rightarrow d}$, (a) magnitude,
(b) phase.\label{fig:hist3_Rsp}%
}

\end{figure}

\begin{figure}
\includegraphics[scale=0.7]{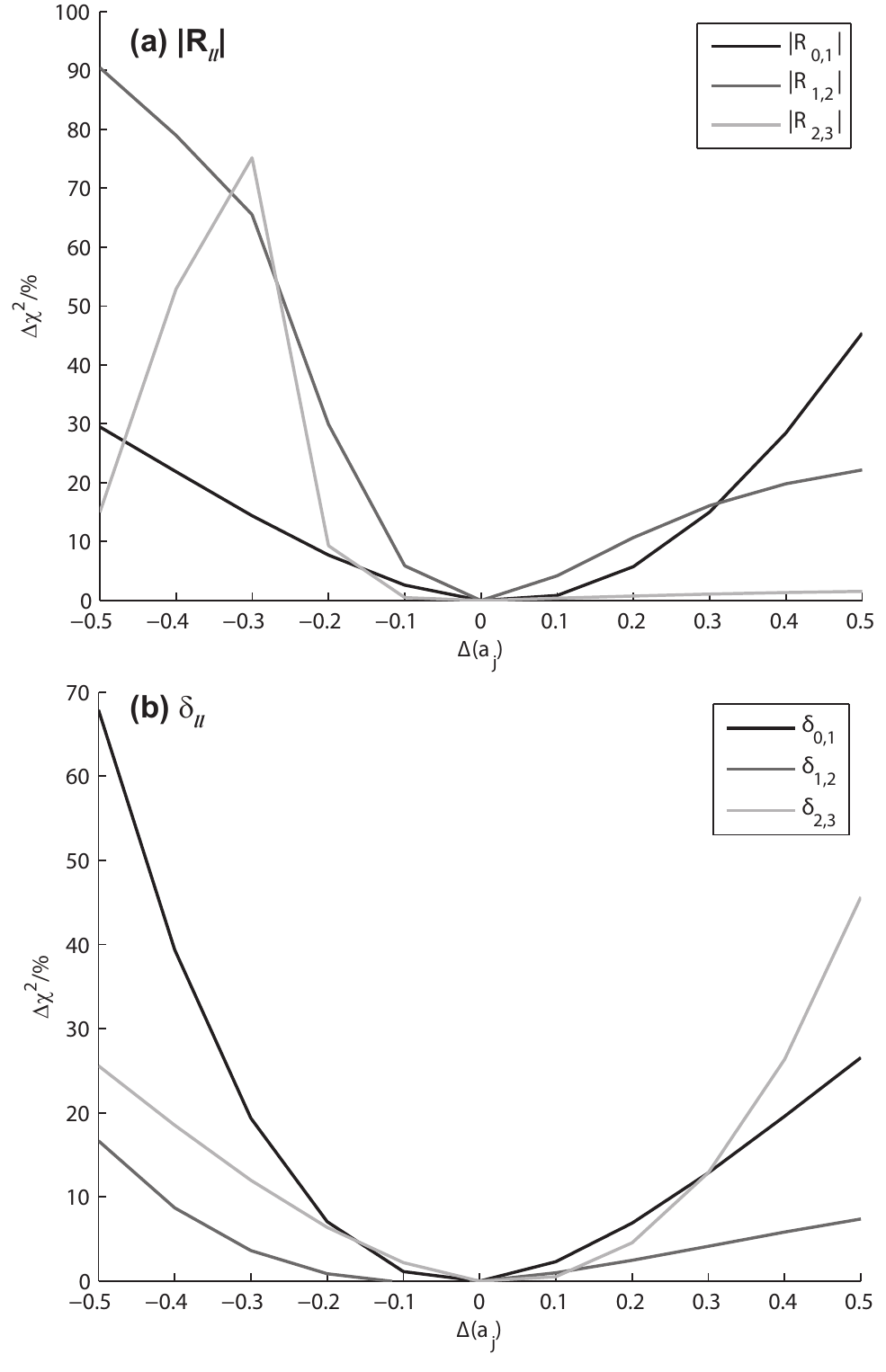}

\caption{%
1D cuts through the $\chi^{2}$ hypersurface for (a) magnitudes and
(b) phases of the $R_{i\rightarrow f}$ matrix elements. Cuts are
made by varying each parameter by $\Delta a_{j}$, here given as a
fractional variation from the best fit value (i.e. test value $a'_{j}=a_{j}+(a_{j}\times\Delta a_{j})$,
and $a'_{j}=a_{j}$ for $\Delta a_{j}=0$)\textcolor{red}{}, and
the resulting change in $\chi^{2}$ is given as a percentage relative
to the best fit value.\label{fig:1D-chi-cuts}%
}
\end{figure}

\begin{figure*}
\includegraphics{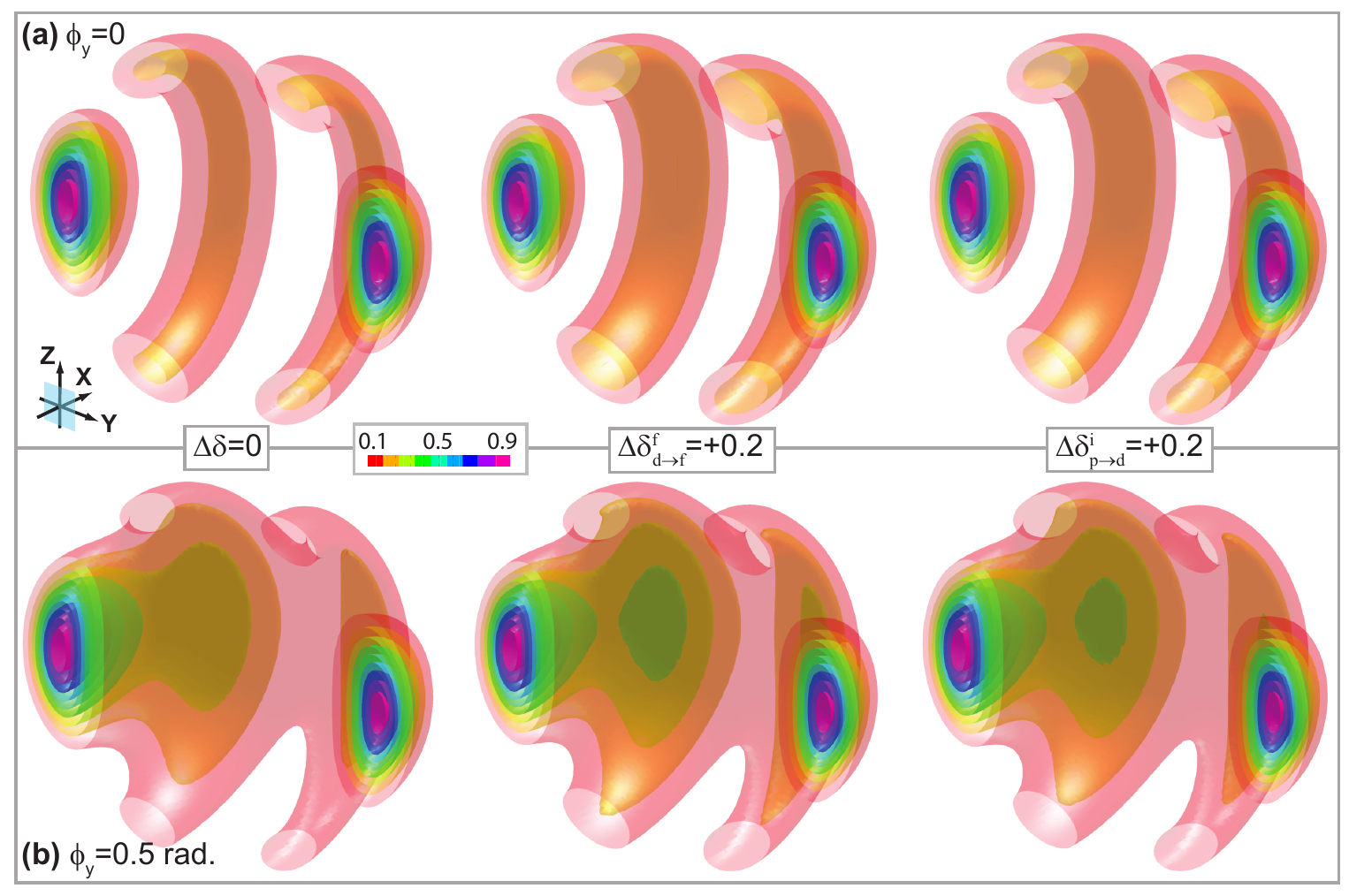}

\caption{%
Calculated 3D distributions $I(\theta,\phi,k)$ (first column) and
sensitivity to changes in the phases $\delta_{d\rightarrow f}^{f}$
and $\delta_{p\rightarrow d}^{i}$ for (a) linear polarization and
(b) elliptical polarization $\phi_{y}=0.5$~rad.\label{fig:Calculated-3D-distributions}%
}

\end{figure*}

As discussed above, within the framework developed herein the radial
matrix elements are the only unknown quantities. With a sufficient
experimental dataset one can therefore hope to obtain these matrix
elements via a fit to the data. In this case the results of such a
fit have already been presented in ref. \cite{Hockett2014}, and validated
via good agreement with both the original 2D imaging data, and additional
3D data obtained via tomographic imaging experiments. We discuss here
further details of the fitting methodology applied, since in general
it is necessary to approach this complicated problem carefully. In
particular, we applied statistical analysis methodologies which were
previously developed for energy-domain photoionization experiments
\cite{Hockett2009,hockettThesis}.

In our procedure, the data from 2D measurements was compared with
the calculated 2D images, as illustrated in figure \ref{fig:computed-PADs}
and obtained as detailed above. The calculated images were then optimized
via a fitting routine, with the radial matrix elements and image generation
parameters as the free parameters for fitting. The criteria for the
best fit was simply the minimization of the sum of least squares:

\begin{equation}
\chi^{2}=\sum_{\theta_{2D},k}(I_{2D}(\theta_{2D},k)-I_{Expt.}(\theta_{2D},k))^{2}
\end{equation}
where $I_{2D}$ is the calculated distribution defined in eqn. \ref{eq:I2D},
and $I_{Expt.}$ the 2D experimental data. This methodology is completely
general, and only relies on the underlying theoretical framework correctly
describing the physics inherent to the problem. However, the size
of the $\chi^{2}$ hyperspace may be very large since it has dimensions
equal to the number of free fitting parameters. The practical outcome
of this is that the possibility of local minima in the hyperspace
is significant, and the parameters obtained via such a procedure must
be carefully evaluated and tested to confirm their veracity and robustness.

In this particular case, the full calculation required 12 parameters,
consisting of the amplitudes and phases of the 5 radial matrix elements
$R_{ll}$ and 2 image generation parameters (Gaussian centre and FWHM)
\footnote{In this case parameters to allow for rotation of the image in the
detector plane were not included, but in general could also be included.%
}. Since absolute phases cannot be determined, one phase is chosen
to be a reference and set to zero, leaving 11 free fit parameters.
Furthermore, the image generation parameters do not have a large influence
on the final results, which are primarily sensitive to the angular
coordinate, and could therefore be bounded quite tightly  after some
initial by eye optimization, thereby reducing the search-space of
physical relevance to, effectively, 9 dimensions. In the fitting procedure
the $R_{ll}$ were expressed in magnitude and phase form, $R_{ll}=|R_{ll}|e^{i\delta_{ll}}$,
where $0\leq R_{ll}\leq1$, $-\pi\leq\delta_{ll}\leq\pi$, and $R_{l_{1}\rightarrow l_{2}}=R_{l_{2}\rightarrow l_{1}}^{*}$.
Fitting was implemented with a standard fitting algorithm, Matlab's
\noun{lsqcurvefit}, based on a Trust-Region-Reflective least-squares
method. The dataset for fitting consisted of four experimental images,
each corresponding to a different pure polarization state of the light,
similar to the states shown in figure \ref{fig:Calculated-3D-distributions}.
The laser pulse was modelled with $\tau=30$~fs and four polarization
states given by $\phi_{y}=0$ (linear polarization, ellipticity $\varepsilon=0$),
$\phi_{y}=\pi/8,\,\pi/4$ (elliptical polarization states, with ellipticities
$\varepsilon\sim$0.2,~0.4) and $\phi_{y}=\pi/2$ (circular polarization,
$\varepsilon=1$). The ellipticities given here are defined as the
ratio of the minor to major axes of the polarization ellipse, hence
$\varepsilon=0$ for linearly polarized light and $\varepsilon=1$
for pure circularly polarized light. Because the elliptical polarization
states may be slightly different from those obtained experimentally
(via the use of a quarter-wave plate, see ref. \cite{Hockett2014}
and refs. therein for further experimental details) the subsequent
fitting was weighted towards the linear and circular polarization
results by an additional factor of two in $\chi^{2}$.

In order to carefully test for local minima the hyperspace was repeatedly
sampled using a Monte-Carlo approach, in which the fitting was repeated
$N$-times with the seed values for the fitting parameters randomized
on each iteration. Statistical analysis of the fitted parameters
derived from such repeated fits can be employed to probe the behaviour
of the fitting algorithm, and also to gain information on how well
the experimental data defines each fitted parameter. Although it is
non-trivial to visualize the full $\chi^{2}$ hypersurface, aspects
can be probed by plotting histograms and correlation plots of the
fitted parameters. A large scatter in the value of a given fit parameter
over a range of fits to the same data suggests a poorly defined parameter;
a consistent result meanwhile shows that a particular parameter is
well defined by the dataset. The experimental data can show different
sensitivities to different parameters depending on the type of ionizing
transitions present, because different transitions will (according
to the magnitude of the geometrical parameters and symmetry constraints)
be more sensitive to certain partial-waves. Additionally, the presence
of multiple-minima in the fit may be revealed by the presence of more
than one feature in the histogram, reflecting more than one ``best''
fit result, while correlations appearing between supposedly uncorrelated
parameters can indicate emergent behaviours in the high-dimensional
space or - more prosaically - issues with the fitting methodology
or coding.

In this case we performed 300 fits, and the lowest $\chi^{2}$ was
obtained on 4 of these fits, which we take to be the absolute minimum.
The radial matrix elements ultimately found, as reported in ref. \cite{Hockett2014},
are given in table \ref{tab:Fitted-params} for reference, and discussed
further below. Figure \ref{fig:hist3_Rsp} gives an illustrative example
of the fitting statistics, in this case showing correlation histograms
between $\chi^{2}$ and the (a) magnitude and (b) phase of $R_{p\rightarrow d}$.
The plot shows the 40 fit results within 5\% of the lowest $\chi^{2}$.
Interestingly, in this case many of the fits are bunched, with $\chi^{2}\sim300$
(arb. units). This most likely reflects the presence of local minima
as defined above, but may also be related to the convergence criteria
set on the fitting algorithm which, in this case, was set to a limited
number of iterations in order to cap the computational time per fit
and ensure a large seed-space for the search; in effect the large
seed-space becomes part of the fitting criteria. Depending on the
seed values, the overall convergence of the fit may be fast or slow,
and the possibility of finding the global minima will vary depending
on the start position in the 11-dimensional parameter-space, as well
as the topography of this space and the details of the fitting algorithm.
From the histograms of the bunched results, it is apparent that $|R_{p\rightarrow d}|$
is somewhat well-defined at the larger $\chi^{2}$, with values mostly
close to the best result, while the phase appears much less well-defined
at this level. At lower $\chi^{2}$, the parameter-space is much sparser,
with only a few parameter sets found, but they appear to converge
on a single parameter set. These observations illustrate the difficulty
in assessing best fit results without careful analysis: in this case
sampling of only a few fit results would potentially lead to a parameter
set quite different from the global optimal found. Here statistical
analysis, as well as further validation of the results against additional
experimental data (see sect. \ref{sec:Photoionization-matrix-elements}),
both serve to provide confidence that the absolute best fit, hence
physically correct, results have been obtained.

\begin{table}
\begin{centering}
\begin{tabular}{c|c|c|c|c|c}
\multicolumn{3}{c|}{Transition} & $|R_{l_{1}l_{2}}|$  & $|R_{l_{1}l_{2}}|^{2}$/\%  & $\delta_{l_{1}l_{2}}$/rad.\tabularnewline
\multicolumn{1}{c}{%
} & \multicolumn{1}{c|}{$l_{1}$} & $l_{2}$  & %
 & %
 & %
\tabularnewline
\hline 
\hline 
$i\rightarrow v$  & p  & s  & 0.34 (3)  & 12 (4)  & 0 {*}\tabularnewline
\hline 
 & p  & d  & 0.94 (8)  & 88 (11)  & -1.62 (4)\tabularnewline
\hline 
$v\rightarrow f$  & s  & p  & 0.85 (8)  & 72 (12)  & -0.19 (3)\tabularnewline
\hline 
 & d  & p  & 0.14 (2)  & 2 (2)  & -2.08 (8)\tabularnewline
\hline 
 & d  & f  & 0.51 (9)  & 26 (13)  & 0.24 (7)\tabularnewline
\end{tabular}
\par\end{centering}

\caption{Fitted values for the \emph{relative} transition matrix element magnitudes,
$|R_{ll}|$, and phases, $\delta_{ll}$. The square of the magnitudes
is expressed as a percentage of the total transition amplitude, normalized
to unity for each step. Uncertainties in the last digit are given
in parentheses. {*} reference phase, set to zero during fitting.\label{tab:Fitted-params} }
\end{table}

\subsection{Robustness, uncertainties \& validation}

 As well as statistically evaluating fit results, the behaviour of
$\chi^{2}$ can also be more directly probed. In essence, this amounts
to removing the black-box nature of the fitting algorithm by explicitly
looking at the gradient and curvature of $\chi^{2}$ as a function
of the fitting parameters, rather than looking at only the final fitted
results. Additionally, the curvature with respect to a given parameter
can be used to provide uncertainty estimates on the fitted parameters
\cite{bevington}:

\begin{equation}
\sigma_{j}^{2}=2\left(\frac{\partial^{2}\chi^{2}}{\partial a_{j}^{2}}\right)^{-1}\label{eq:sigma}
\end{equation}
where $\sigma_{j}$ is the uncertainty in parameter $a_{j}$. Equation
\ref{eq:sigma} relates the response of $\chi^{2}$ to a given parameter;
the sharper the response the better $a_{j}$ is defined by the data
and hence the smaller the uncertainty. In practice this procedure
equates to varying each fitted parameter by $\Delta a_{j}$, and evaluating
$\chi^{2}$ for this new parameter set, in order to map out 1D cuts
through the $\chi^{2}$ hypersurface. Uncertainties estimated in this
manner were given in ref. \cite{Hockett2014}, and are provided again
in table \ref{tab:Fitted-params}. It is also of note that a similar,
but not identical, procedure can be performed by refitting all other
$(n-1)$ parameters as a function of a test parameter $a_{j}$ \cite{Gessner2002}.
This procedure will also provide 1D cuts through the hypersurface,
but along the $n$-dimensional topography of the minimum. The drawback
of this alternative procedure is the necessity of performing many
additional fits, which may be computationally expensive; for this
reason it was not explored in this work.

Figure \ref{fig:1D-chi-cuts} shows 1D cuts through the $\chi^{2}$
hyperspace as defined above, for the magnitudes and phases of $R_{i\rightarrow f}$.
In this case it is clear that the sensitivity of $\chi^{2}$ is good
in most cases, with 10~\% changes in $a_{j}$ (i.e. $\Delta a_{j}=\pm0.1$)
typically leading to clear changes in $\chi^{2}$; this is also reflected
in the relatively small uncertainties $\sigma_{j}^{2}$ given in table
\ref{tab:Fitted-params}. In this case, a notable exception is $|R_{2\rightarrow3}|$,
which is much less sensitive to $\Delta a_{j}$ for increases in magnitude.
This is the magnitude of the $f$-wave channel, which dominates the
ionization overall; consequently the final PAD is not very sensitive
to small increases in the magnitude of this matrix element, although
does remain very sensitive to decreases in magnitude, and its relative
phase. In general $\chi^{2}$ is somewhat less sensitive to the phases
than the magnitudes, although the response is still significant. It
is also of note that the 1D cuts are not symmetric about $\Delta a_{j}$,
reflecting the complicated topography of the $\chi^{2}$ hypersurface,
and the fact that it is dependent on the relative, rather than absolute,
values of the matrix elements.

A final, valuable test of the determined matrix elements is their
predictive power, and the possibility of testing such predections
against additional experimental results not used in the original extraction
procedure. A consideration of the sensitivity of the determined matrix
elements in these terms is a useful way of evaluating the results.
In previous, energy domain, studies the (rotational) energy spectrum
could be used to provide the additional, independent data against
which the extracted matrix elements could be further verified \cite{Hockett2009},
and the possibility of using different polarization geometries combined
with tomographically reconstructed PADs was also explored \cite{Hockett2010}.
As noted above, and discussed briefly in ref. \cite{Hockett2014},
comparison of the current results with 3D photoelectron data obtained
via tomography was also employed in this case. The comparison with
the experimental data is discussed in sect. \ref{sub:Quantitative-comparison},
while the sensitivity of the computed 3D distributions to changes
in the matrix elements $\Delta a_{j}$ is discussed here. Figure \ref{fig:Calculated-3D-distributions}
provides some examples of this sensitivity for the variation of two
different phases by 20~\%, and for two different polarizations. Although
the sensitivity of the 3D distributions to these phases is inherent
in the small uncertainties determined above, as well as the ability
to successfully use a fitting methodology, it is nonetheless instructive
to visualize the sensitivity in this way. Here it is clear that, while
both cases exhibit a sensitivity to the phase adjustments, the changes
in the linearly polarized case are less significant. In this case,
the width of the central bands increases slightly and, although this
change still correlates with a change in the $\beta_{L,M}^{k}$, the
magnitude of this change means that it will only be revealed by careful
quantitative analysis, and may not be obvious in a qualitative comparison.
This conclusion becomes even stronger for 2D images of this distribution.
In the elliptically polarized case the phase changes are manifested
in an increase in flux and spread of the equatorial lobes of the distribution,
and are much more pronounced as compared to the linearly polarized
case. While the sensitivity observed here merely confirms the earlier
analysis, the investigation of the predicted distributions in this
phenomenological manner provides additional insight into the fitting
process, in particular the magnitude of changes which might be expected
in a given case and, hence, suggests possibilities for future experimental
work, particularly in the more complex case of shaped pulses, as discussed
in ref. \cite{Hockett2014}.

Overall the methodology outlined here might be viewed as a pragmatic
approach to complete experiments. Utilizing a combination of fitting,
Monte-Carlo sampling, direct exploration of the $\chi^{2}$ hyperspace
and further validation of the results based on their predictive power,
a careful validation of their robustness and validity can be made
for the case at hand. This is distinct from a more formal treatment,
such as that discussed in Schmidtke et. al. \cite{Schmidtke2000},
wherein the fundamental limits of a fitting approach are derived.
In the current work a comparison with the definitions given in that
work have not been made, but the pragmatic methodology herein indicates
that the extraction of the matrix elements is, in this case, reliable.
An extension of this methodology, combined with a formal treatment,
to investigate the additional possibilities in the polarization-multiplexed
case remains for future work.

\section{Comparisons with tomographic data\label{sec:Photoionization-matrix-elements}}

Here we focus on a detailed comparison of the results of ref. \cite{Hockett2014}
with additional experiments which provided full 3D data. These results,
obtained using photoelectron tomography techniques (see ref. \cite{Hockett2015b}
for details), provide both a highly detailed volumetric data and a
set of measurements at a different laser intensity. The former characteristic
allows for a qualitative visual comparison of 3D distributions, which
reveal details of the distributions which may be obscured in the 2D
images, and the possibility of a full retrieval of the $\beta_{LM}(k)$
from the data, which is not possible for non-cylindrically symmetric
2D images and allows for a more quantitative comparison of experiment
and theory. The use of different intensities ($\sim10^{13}$~Wcm$^{-2}$
for the tomographic data, as compared to $\sim10^{12}$~Wcm$^{-2}$
for the 2D data) provides further evidence for the lack of any significant
strong-field effects to the angular distributions in this case, and
the veracity of our ionization model.

\subsection{Qualitative comparison}

\begin{figure*}
\includegraphics[scale=1.1]{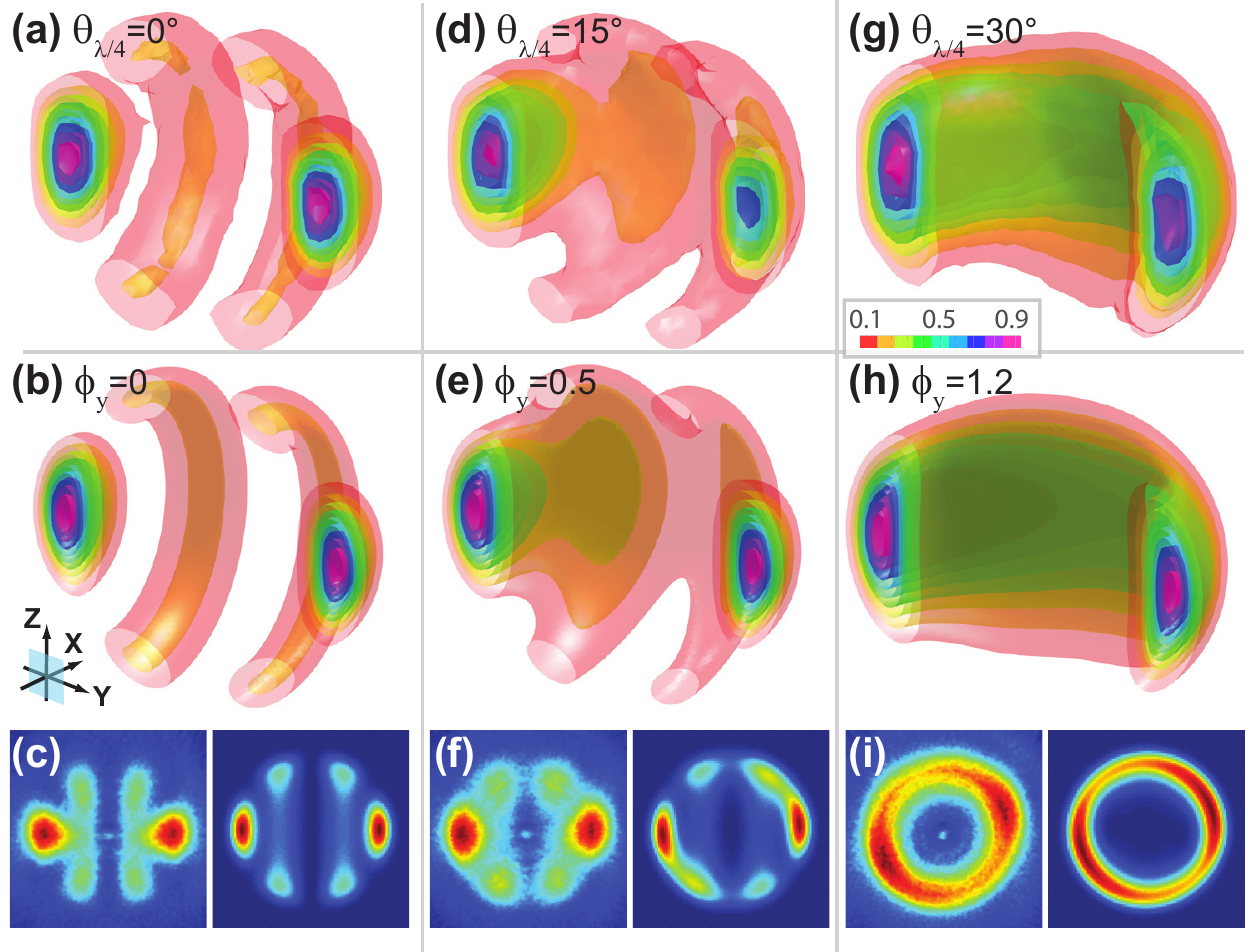}

\caption{%
Comparison of calculated \& experimental tomographic results. Panels
show (top) experimental and (middle) calculated 3D distributions plotted
as intensity iso-surfaces. The experimental data is plotted with a
radial mask to show only the main part of the radial distribution.
(Bottom) Experimental and calculated 2D projected images in the plane
of polarization $(X,Y)$. In this case no radial mask is included
in the experimental results, and the images show an extended energy
range relative to the computational results (the broader spectrum
is due to the presence of Autler-Townes splitting in this case, see
main text for details). Columns show the results for different laser
polarization states, defined by a quarter-wave plate rotation angle
($\theta_{\lambda/4}$) in the experimental data, and spectral phase
$\phi_{y}$ in the calculations. The states correspond to ellipticities
$\varepsilon\backsimeq$0,~0.3 and 0.6 (\foreignlanguage{british}{see
}ref. \cite{Hockett2015b}\foreignlanguage{british}{ for further detail}s).\label{fig:tomo-qualitative}%
}

\end{figure*}

A qualitative comparison of data for the three different polarization
states measured is presented in figure \ref{fig:tomo-qualitative}.
In this figure the full 3D distributions are shown as nested iso-surface
plots, and 2D images in the polarization plane are also shown. In
this case, the experimental data has an additional high-energy feature
in the radial distribution, arising from Autler-Townes splitting which
becomes significant at higher intensities (see refs. \cite{Wollenhaupt2005b,Wollenhaupt2009a},
and ref. \cite{Hockett2015b}). In this analysis only the main feature
is of interest, and the tomographic distributions shown include a
radial mask in order to remove the additional contributions and facilitate
comparison over the main spectral feature. For the 2D images (bottom
row of fig. \ref{fig:tomo-qualitative}) no radial mask is employed
and, consequently, the experimental results show a broadening of the
spectrum in the 2D images. Full details of the experimental data and
tomographic reconstruction procedure, as well as the energy spectra,
are discussed in ref. \cite{Hockett2015b}. It is also of note that
the data shown in panels (a) and (d) of figure \ref{fig:tomo-qualitative}
are the same as shown in figure 3 of ref. \cite{Hockett2014}.

It is clear from figure \ref{fig:tomo-qualitative} that the experimental
data and the calculations agree in overall form, with the trend in
the shape of the PADs with polarization well-reproduced by theory.
This general behaviour is not surprising, since this sensitivity was
inherent in the concept of obtaining the photoionization matrix elements
via a fitting procedure from the 2D images recorded with different
polarizations, but does indicate that the additional details observable
in the tomographic data do not contradict the fit results, even at
higher intensities which do affect the energy spectrum.

The 2D images appear to show less satisfactory agreement but, since
these $(X,Y)$ plane projections include summation over the $Z$-axis,
this is perhaps unsurprising. In particular, the apparent increase
in intensity of the band structures in panel (c), relative to the
computational results, is due to the additional (and incoherent) contribution
from photoelectrons at different energies, due to the projection of
the broader spectrum onto the 2D plane, and which are not present
in the computational results. The most significant differences are
seen in panel (f), where the asymmetry in the $(X,Y)$ plane - the
helicity of the distribution - is reduced relative to the computational
results. This is likely due to a slight difference of the polarization
ellipse relative to the calculations, as well as the summation over
the broader spectrum (as mentioned above) which may wash-out fine
details in the projection image. For the results approaching circular
polarization, panel (i), the agreement is better. In this case the
contributions from the higher-energy AT feature are reduced (see ref.
\cite{Hockett2015b}), and the polarization state of the light may
be slightly better matched to that assumed in the calculation.

Overall these results indicate reasonably good agreement between the
previously determined matrix elements and the tomographic data, but
also indicate the problematic aspects of a qualitative comparison
for these complex distributions. In general such comparisons are worthwhile,
but subject to perceptual bias which may be highly dependent on the
type of data visualization used. Naturally a quantitative comparison
is preferable, and is explored in the following section.

\subsection{Quantitative comparison\label{sub:Quantitative-comparison}}

\begin{figure}
\includegraphics{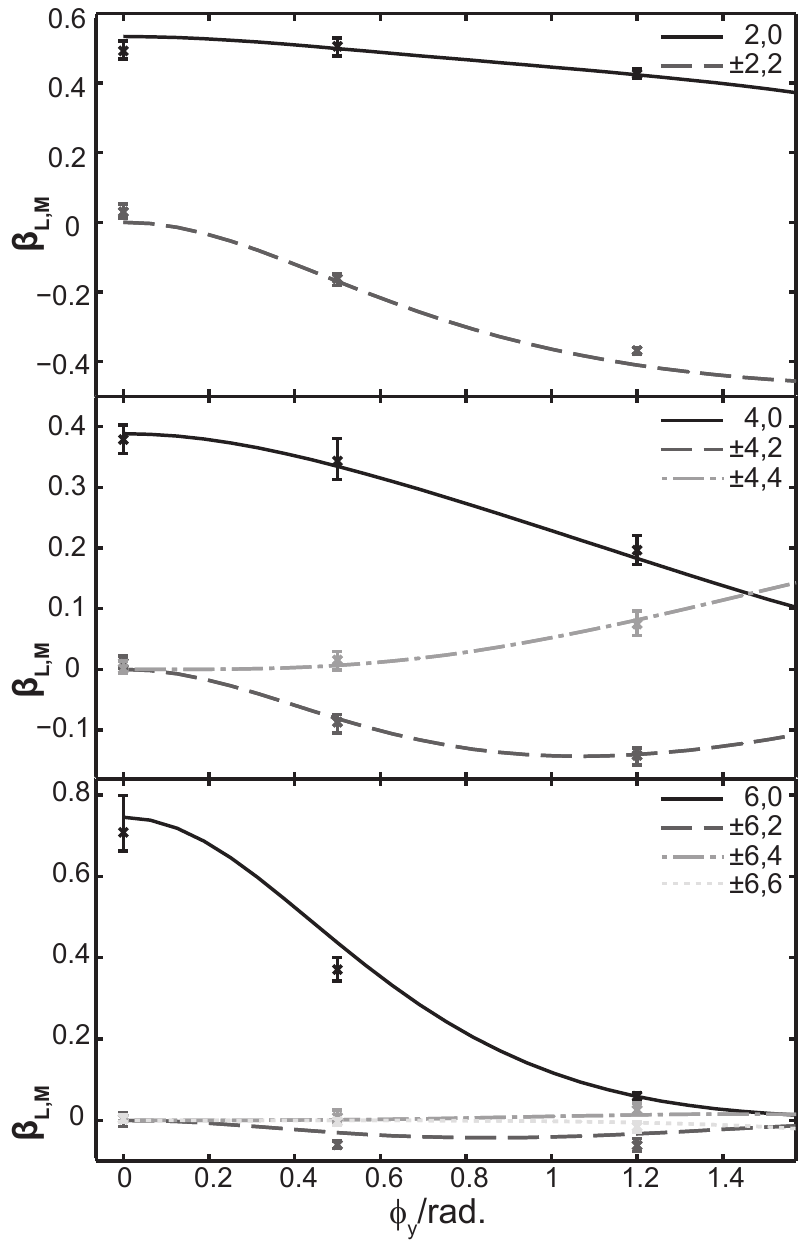}

\caption{%
Comparison of calculated $\beta_{LM}^{k}$ (lines) and experimental
results (crosses with error bars) as a function of polarization state.
Panels show $L=2,\,4,\,6$, and line-styles and shade denote $M$.
The error bars on the experimental results show the spread of $\beta_{L,M}(k)$
obtained over the FWHM of the main spectral feature.\label{fig:BLM_calc_vs_expt}%
}
\end{figure}

To make a more careful comparison of the volumetric results, $\beta_{L,M}(k)$
were extracted from the data (see ref. \cite{Hockett2015b} for details).
The $\beta_{L,M}(k)$ over the main feature can then be directly compared
with the predicted $\beta_{L,M}^{k}$ based on the fitted matrix elements.
Since the fit results assume that the matrix elements are approximately
constant over the feature, the experimental $\beta_{L,M}(k)$ were
averaged over the FWHM of the main spectral feature to yield an energy-averaged
value. These values are plotted in figure \ref{fig:BLM_calc_vs_expt}
along with the calculation results. The range of the experimental
$\beta_{L,M}(k)$ are indicated by the error bars on the plot, indicating
the spread of values over the spectrum.

The agreement between the calculation and experimental results is
generally very good, if not exact. The dominant terms, with $L=2,\,4,\,6$
and $M=0$, show excellent agreement and, aside from $\beta_{6,0}$
at $\phi_{y}=0$, the experimental results also show only a small
spread of values. For the $M\ne0$ terms, which generally have smaller
magnitudes than the $M=0$ terms except at large $\phi_{y}$, the
agreement is generally good, but less so for the $|M|=2$ terms. Here
the trends with polarization are in agreement, but the exact values
are shifted slightly from the calculations. As noted above, these
small discrepancies may be due to slight differences in the laser
polarization and frame rotations used in the calculations as compared
with the experiment.

Additionally, the experimental data indicated a significant energy-dependence
of the PADs away from the main spectral feature, small contributions
from higher-order terms ($L>6$) and symmetry breaking were also present.
These effects are not accounted for by the net 3-photon model, and
indicate the presence of additional complexities to the light-matter
interaction. These additional effects are, at this time, not well-understood
beyond the clear requirement for $\pm m$-state symmetry breaking,
and for higher angular momentum states to be accessed. These observations
are discussed further in ref. \cite{Hockett2015b}, and it is of note
that the ability to resolve these additional effects via the quantitative
analysis of 3D photoelectron data is a significant outcome. 

Despite these additional, but intriguing, complexities, the major
channels observed over the FWHM of the main spectral feature are seen
to agree very well with the previous analysis, based on 2D data recorded
at lower intensities, overall providing a strong test of the accuracy
of the ionization matrix elements determined in that case. The possibility
of gaining a detailed understanding of the additional effects observed
in the 3D data, starting from the current 3-photon model, and associated
ionization matrix elements, remains an interesting proposition for
future work. In the following section we explore some extensions to
our treatment which may facilitate such understanding.

\section{Assumptions, extensions \& physical considerations\label{sub:Assumptions-&-extensions}}

In the above treatment, as applied in ref. \cite{Hockett2014}, some
simplifications have been made for the specific case at hand, in order
to facilitate the determination of the $R_{ll}(k)$ as detailed above
(sect. \ref{sec:Photoelectron-image-generation}). Here, in order
to generalize this treatment further, we consider more carefully the
assumptions made and explore other extensions to the theory.

\subsection{Atoms}

The intra-pulse dynamics above implicitly assume that only the outermost
electrons play a role in the intra-pulse dynamics, and that lower-lying
bound states can be neglected. Furthermore, it assumes that the 4$p$
manifold is the only unpopulated state which plays a role at the 1-photon
level. This is expected in this case because the 4$s\rightarrow$4$p$
transition carries significant oscillator strength, and is near resonant
with the laser pulse. However, in general it is possible that other
states will play a role, particularly as the laser is tuned further
from the 4$s\rightarrow$4$p$ line. This would result in a more complex
TDSE (eqn. \ref{eq:TDSE}), with additional states appearing, and
also necessitate a more careful treatment of the transition dipoles
$\mu_{L/R}$ to allow for variation in the transition amplitudes to
different $|n,l\rangle$-manifolds. The treatment of the 2-photon
ionization would, similarly, increase in complexity with the addition
of further initial/source $|n,l\rangle$-manifolds, but would otherwise
remain identical. 

In the case of atomic ionization with a structureless continuum, the
photoelectron energy spectrum can be treated somewhat directly as
determined from the power spectral density of the laser pulse \cite{Wollenhaupt2002,Wollenhaupt2005,Wollenhaupt2009a}.
Such treatment effectively introduces an additional time-dependent
phase into eqn. \ref{eq:d_if-kt}. In ref. \cite{Wollenhaupt2009a}
this phase is defined as $e^{i\delta\omega_{e}t}$, with $\delta\omega_{e}=\omega_{e}+\omega_{IP}-\omega_{p}-2\omega_{0}$,
where the angular frequencies are related to the electron energy ($\hbar\omega_{e}$),
the ionization potential ($\hbar\omega_{IP}$), the ionizing 4$p$
state ($\hbar\omega_{p}$) and the photon energy ($\hbar\omega_{0}$).
This phase will thus oscillate rapidly at the resultant difference
frequency of these terms (effectively the difference between the total
final state energy and the incident/input energy), and directly gives
rise to a photoelectron energy spectrum dependent on the pulse properties,
including its temporal duration and structure \cite{Wollenhaupt2006}.
This dependence can be considered interferometrically, in the sense
that the resultant (time-integrated) energy spectrum is the coherent
temporal sum, hence contains interferences between all instantaneous
momentum distributions; this is exactly analogous to the PADs considered
as the coherent temporal sum of the instantaneous angular distributions
(at a given energy). In the case of pulses intense enough to create
significant Autler-Townes splitting in the photoelectron energy spectrum,
this treatment could allow for a description of the changes in the
PADs and symmetry breaking, as discussed in sect. \ref{sub:Quantitative-comparison}
and correlated with the Autler-Townes doublet in the spectrum. This
consideration is discussed further in ref. \cite{Hockett2015b}.

In the most general case, where multiple, non-degenerate ionization
pathways may be present, interferences may arise between ionizing
transitions with very different angular structures. In the energy
domain this effect has been investigated by Elliott and co-workers
in experiments utilizing fundamental and second-harmonic light to
create final-state interferences between different intermediates \cite{Yin1992,Yin1995,Wang2001}.
Control over the relative phase of the two colours allowed for control
over the resultant interferences \cite{Yin1995}. A similar concept
was also employed to measure the phase of a bound-state \cite{Fiss2000}.
Practically, this most general effect could be included in our formalism
by the inclusion of sets of ionization matrix elements correlated
with the distinct sets of ionization pathways, where each set has
a characteristic partial wave distribution (amplitudes and phases)
and energy-dependent phase factor, and would result in the inclusion
of interferences dependent on both geometric and energetic phase factors.
Conceptually, this effect is inherent in the PADs arising from polarization
shaped pulses, where the different ionization pathways correspond
to different intermediate angular momentum states, but in this case
all levels are degenerate and the relevant phase shifts are purely
geometric.

\subsection{Molecules}

In the case of molecular ionization the situation is more complex.
In this case, the partitioning of the incident photon angular momentum
to molecular rotations, as well as the outgoing photoelectron, requires
a more involved treatment of the geometric terms, even in a single
active electron picture. Furthermore, one might expect that the continuum
also contains structure due to population of different vibrational
modes of the ion, although it is also possible that these states have
little effect on the $R(k)$ integrals over a small energy range.
This assumption formally means a separation of the ionization matrix
element into electronic, vibrational (Franck-Condon) and rotational
terms is possible, and that these terms are thus uncoupled. Effectively
the electronic terms define the $R(k)$, and the Franck-Condon factors
an overall transition intensity envelope - but one that does not affect
the partial wave character of the continuum. In some cases this approximation
has been tested, and found to hold, but in other cases - particularly
when considering highly excited vibrational modes - one might expect
this assumption to fail \cite{Lucchese2010,Lopez-Dominguez2012}.

In previous work we have investigated molecular ionization via vibrational
and rotational state-resolved energy-domain experiments \cite{Hockett2007,Hockett2009,Hockett2010,Hockett2010a}.
This work demonstrates the feasibility of performing such experiments,
and illustrates the types of angular momentum coupling schemes required.
Although the current work, incorporating intra-pulse dynamics and
polarization shaped pulses, has not yet been extended to molecular
cases, such an extension seems feasible based on these earlier studies,
at least from the perspective of treating the ionization matrix elements,
and including the larger number of continuum $l$-waves required for
molecular scattering problems.

The intra-pulse dynamics in the molecular case may, however, be significantly
more challenging. Clearly there are many more degrees of freedom to
account for, and the potential for both nuclear and electronic wavepacket
motion during the pulse, as well as the possibility of dumping a lot
of angular momentum into forming a rotational wavepacket during a
strongly-coupled initial step (although the evolution of this wavepacket
will ultimately be on a much slower timescale). The incorporation
of such coupled rovibronic dynamics is, in practice, quite difficult
due to the high-dimensionality of the problem. It is certainly not
sufficient to perform a simple TDSE of coupled electronic states as
employed herein, although the coupling of more complex wavepacket
calculations with the ionization treatment herein would be feasible.
A conceptually similar, although fully \emph{ab initio}, coupling
of complex vibronic wavepackets with a full photoionization calculation
has recently been presented for the triatomic molecule CS$_{2}$ \cite{Wang2014};
prior to this \emph{ab initio} treatment a simpler dynamical model
was combined with the relevant angular momentum coupling and gave
good agreement with experimental results, although the treatment was
only semi-quantitative and stopped short of extraction of the ionization
matrix elements \cite{Hockett2011}. For diatomics it is probable
that a conceptual middle-ground is found, in which the required low-dimensionality
wavepacket can be modelled via a simple TDSE treatment with enough
accuracy to be of use. For polyatomics, the complexity of the wavepacket
will be the deciding factor, depending directly on the number and
type of states and couplings involved in a given case. In the most
complex cases the problem may be best treated by fully \emph{ab initio}
calculations including photoionization, the results of which can be
compared directly with experimental data at a high level, but in simpler
wavepackets (few level and/or weakly coupled) a basic TDSE approach
may be of sufficient accuracy to be useful for complete experiments.

In sum, based on experience of similar problems in molecular ionization
in both the energy and time domains, it seems feasible that this time-domain
multiplexing concept, employing multi-photon ionization schemes, and
with the inclusion of intra-pulse dynamics, can also be applied successfully
to (at least some) molecular photoionization problems.

\subsection{Other regimes}

Other regimes are also of general interest in photoionization studies
\cite{Reid2003}, in particular the strong-field (non-perturbative)
regime. Very generally, the treatment presented herein could be extended
to this regime, and the issues associated closely follow the discussion
above. In the non-perturbative case the intra-pulse dynamics become
more complicated, since a single-active electron picture is no longer
likely to be valid, and a static picture of the bound state energy-level
structure also breaks down. Similarly, the scattering dynamics of
the outgoing electron will also be time-dependent, since the scattering
must now incorporate the laser-induced part of the potential, not
just the (static) atomic or molecular potential. 

In theory it is feasible to allow for these effects into the treatment
presented herein, since it is already time-dependent and, as discussed
for the atomic and molecular cases above, additional dynamical effects
could be readily incorporated providing the numerics are tractable
and accurate. However, the main issue in terms of determining the
ionization matrix elements would be the large size of the set of matrix
elements to be determined in a fully time-dependent treatment and
the concomitant complexity of the fitting procedure if a set of ionization
matrix elements were required for each time-step. In such cases it
may be possible to posit an effective functional dependence of the
ionization dynamics on the laser field to mitigate this somewhat,
but one would have to more carefully consider exactly what kind of
measurement would allow for a unique set of (fitted) matrix elements
to be extracted from the (necessarily) time-integrated photoelectron
image. 

Another regime is that of high-order light-matter couplings beyond
the dipole approximation. In this case the ionization matrix elements
contain higher-order angular momentum couplings, hence a more complex
angular-momentum coupling scheme and a larger set of matrix elements
must be determined, similar to the considerations for the molecular
case. As for that case, there is no fundamental reason why such cases
could not be treated within the theoretical framework presented here,
although the feasibility of the fitting procedure would have to be
assessed for any given case based on the size of the problem.

Another interesting extension is to ionization time delays, since
the the Wigner delay time is given by the energy-derivative of the
scattering phase \cite{Wigner1955,DeCarvalho2002}. Measurements of
this phase, based on the concept of interfering photoelectron wavepackets
created with different energies, have recently been demonstrated \cite{Klunder2011,Dahlstrom2012}
and are very similar to to concepts herein. In such measurements,
above-threshold ionization creates electron wavepackets at different
energies (i.e. multiple spectral features in the photoelectron spectrum),
and further photo-absorption from a probe laser field can be used
to interfere neighbouring wavepackets. This procedure results in side-band
generation in the photoelectron spectrum, and the phase of the oscillation
of these side-bands with respect to the probe field timing provides
information on the relative phase of the photoelectron wavepackets
- this is know as the RABBITT technique \cite{Muller2002,Dahlstrom2012}
(this concept is somewhat analogous to the two-pulse photoelectron
interferometry of ref. \cite{Wollenhaupt2002}). The difference between
this concept and traditional ``complete'' photoionization experiments
is that the total photoionization phase is measured over a broad energy
spectrum in RABBITT measurements, as opposed to the measurement of
the phases of the partial-waves at a single energy as discussed herein.
By extending our technique to a broad energy range, e.g. via observation
of multiple ATI features or the use of a broader bandwidth probe pulse,
we would be able to obtain the partial-wave phases as a function of
energy and, thus, determine the Wigner delay. Furthermore, by obtaining
the phases for all partial-waves, the angle-dependence of the Wigner
delay in molecular ionization could also be investigated \cite{Hockett2015c}.

\subsection{Maximum information measurements \& multiplexing}

In all cases discussed above, the main consideration is the feasibility
of performing complete experiments for more complex, dynamical ionization
schemes. Such applications will, naturally, be challenging, and require
both a detailed theoretical understanding of the dynamics at hand
and high-information experimental measurements. The majority of this
work has focussed on assessing the results obtained for pure polarization
states, in which there is no additional information gained from the
coherent time-domain integration over the laser pulse, but for more
complex cases the additional information content of polarization-multiplex
measurements may be vital. Specifically, multiplexing provides additional
time-domain interferences in the PADs (see eqn. \ref{eq:dtInt}),
with the result that time-integrated polarization-multiplexed measurements
contain the information of multiple pure-state measurements. 

One particularly powerful aspect of using shaped pulses is the possibility
of tailor-made pulses for metrology, designed to create or amplify
specific interfering channels of interest. Conceptually this is identical
to the use of shaped pulses for control \cite{Wollenhaupt2005,Wollenhaupt2009a},
however the pulses would be designed for the purposes of obtaining
detailed information on specific ionization channels, rather than
for the purposes of creating a specific photoelectron distribution.
The examples shown in figure \ref{fig:Time-dependent-dynamics-elliptical}
and \ref{fig:Time-dependent-dynamics-shaped} indicate how this concept
operates: there are different continuum populations created in the
two cases, the time-domain structure is more complex in the shaped
pulse case and, most generally, the pulse shape can be chosen to select
certain ionization channels (within the constraints imposed by the
dynamics of the ionizing system). As well as polarization-shaping,
the coherent time-domain treatment may also provide a way to probe
additional intereferences due to effects such as intensity-dependent
ionization phases. The presence of such effects at higher intensities
has been determined from the behaviour of the PADs over the Autler-Townes
structure of the photoelectron spectrum, as mentioned above (sect.
\ref{sub:Quantitative-comparison}), but remains to be understood
in detail.

In all these cases, the PADs will usually be non-cyclindrically symmetric,
so use of ``maximum information measurements'' utilizing 3D measurements
and detailed analysis will also be required. The power of this approach
has been touched on here, and is further explored in ref. \cite{Hockett2015b};
recent work has also considered 3D photoelectron measurements in the
context of photoelectron circular dichroism \cite{Lux2015}.

\section{Summary \& conclusions}

In this work the validity of a fitting approach to complete photoionization
experiments in the multiphoton regime, incorporating intra-pulse dynamics,
as initially reported in ref. \cite{Hockett2014}, has been explored.
The details of the fitting procedure, based on statistical sampling
of the $\chi^{2}$ hyperspace and further testing and validation of
the results, were outlined as a pragmatic fitting methodology. The
results presented in ref. \cite{Hockett2014} were discussed in detail,
and compared both qualitatively and quantitatively with full 3D experimental
photoelectron distributions. Finally, extension of this treatment
to more complex ionization processes was discussed in general terms.

This analysis indicated the validity of the results already presented,
as well as insight into the practicalities of a pragmatic fitting
approach. Although this approach has yet to be tested beyond the use
of pure polarization states, the use of polarization shaped pulses
clearly offers an enhanced photoelectron metrology, with the possibility
of controlling the information content via the pulse shape, as discussed
in sect. \ref{sec:Intrapulse-dynamics-theory} (see also ref. \cite{Hockett2014}).
The use of full 3D experimental measurements is another powerful aid
to maximum information metrology, as indicated herein by comparison
of the computational results with tomographically reconstructed experimental
distributions (see also ref. \cite{Hockett2015b}). In general, we
anticipate that the combination of these tools represents a powerful
methodology for complete photoionization studies, or other research
making use of ionization measurements.

Acknowledgement: Financial support by the State Initiative for the
Development of Scientific and Economic Excellence (LOEWE) in the LOEWE-Focus
ELCH is gratefully acknowledged.

\bibliographystyle{unsrt}
\bibliography{/media/hockettp/StoreM/reports/bibliography/baumert_paper3_151014_problemRefs,/media/hockettp/StoreM/reports/bibliography/baumert_collab_final_noURL_180215,/media/hockettp/StoreM/reports/bibliography/baumert_paper2_040315}

\begin{thebibliography}{10}

\bibitem{Reid2003}
Katharine~L. Reid.
\newblock Photoelectron angular distributions.
\newblock {\em Annual Review of Physical Chemistry}, 54(1):397--424, 2003.

\bibitem{Cherepkov2005}
N.A. Cherepkov.
\newblock Complete experiments in photoionization of atoms and molecules.
\newblock {\em Journal of Electron Spectroscopy and Related Phenomena},
  144-147:1197 -- 1201, 2005.

\bibitem{Dill1976}
D~Dill.
\newblock {Fixed-molecule photoelectron angular distributions}.
\newblock {\em The Journal of Chemical Physics}, 65(3):1130 -- 1133, 1976.

\bibitem{Lambropoulos1973}
P~Lambropoulos.
\newblock {Using polarization effects in multiphoton ionization to measure
  ratios of bound-free matrix elements}.
\newblock {\em Journal of Physics B: Atomic and Molecular Physics},
  6(11):L319--L321, November 1973.

\bibitem{berry1976}
J.~A. Duncanson, M.~P. Strand, A.~Lindg\aa{}rd, and R.~S. Berry.
\newblock Angular distributions of electrons from resonant two-photon
  ionization of sodium.
\newblock {\em Physical Review Letters}, 37(15):987--990, Oct 1976.

\bibitem{Duong1978}
H~T Duong, J~Pinard, and J~L Vialle.
\newblock {Experimental separation and study of the two partial photoionisation
  cross sections $\sigma$ 3p,s and $\sigma$ 3p,d from the 3p state of sodium}.
\newblock {\em Journal of Physics B: Atomic and Molecular Physics},
  11(5):797--803, March 1978.

\bibitem{Hansen1980}
John Hansen, John Duncanson, Ring-Ling Chien, and R.~S. Berry.
\newblock {Angular distributions of photoelectrons from resonant two-photon
  ionization of sodium through the 3p 0\^{}\{2\}P\_\{3/2\} intermediate state}.
\newblock {\em Physical Review A}, 21(1):222--233, January 1980.

\bibitem{Chien1983}
Ring-ling Chien, Oliver Mullins, and R.~Berry.
\newblock {Angular distributions and quantum beats of photoelectrons from
  resonant two-photon ionization of lithium}.
\newblock {\em Physical Review A}, 28(4):2078--2084, October 1983.

\bibitem{Reid1991}
David~J. Leahy, Katharine~L. Reid, and Richard~N. Zare.
\newblock Complete description of two-photon (1+1[script ']) ionization of no
  deduced from rotationally resolved photoelectron angular distributions.
\newblock {\em The Journal of Chemical Physics}, 95(3):1757--1767, 1991.

\bibitem{Reid1992}
Katharine~L. Reid, David~J. Leahy, and Richard~N. Zare.
\newblock Complete description of molecular photoionization from circular
  dichroism of rotationally resolved photoelectron angular distributions.
\newblock {\em Physical Review Letters}, 68(24):3527--3530, Jun 1992.

\bibitem{Suzuki2006}
Toshinori Suzuki.
\newblock {Femtosecond time-resolved photoelectron imaging.}
\newblock {\em Annual review of physical chemistry}, 57:555--92, January 2006.

\bibitem{Hockett2009}
Paul Hockett, Michael Staniforth, Katharine~L. Reid, and Dave Townsend.
\newblock Rotationally resolved photoelectron angular distributions from a
  nonlinear polyatomic molecule.
\newblock {\em Physical Review Letters}, 102(25):253002, Jun 2009.

\bibitem{Suzuki2012}
Yoshi-Ichi Suzuki, Ying Tang, and Toshinori Suzuki.
\newblock {Time-energy mapping of photoelectron angular distribution:
  application to photoionization stereodynamics of nitric oxide.}
\newblock {\em Physical chemistry chemical physics : PCCP}, 14(20):7309--20,
  May 2012.

\bibitem{Gessner2002}
O.~Ge\ss{}ner, Y.~Hikosaka, B.~Zimmermann, A.~Hempelmann, R.~R. Lucchese,
  J.~H.~D. Eland, P.-M. Guyon, and U.~Becker.
\newblock $4\sigma{}-1$ inner valence photoionization dynamics of no derived
  from photoelectron-photoion angular correlations.
\newblock {\em Physical Review Letters}, 88(19):193002, Apr 2002.

\bibitem{Lebech2003}
M.~Lebech, J.~C. Houver, A.~Lafosse, D.~Dowek, C.~Alcaraz, L.~Nahon, and R.~R.
  Lucchese.
\newblock Complete description of linear molecule photoionization achieved by
  vector correlations using the light of a single circular polarization.
\newblock {\em The Journal of Chemical Physics}, 118(21):9653--9663, 2003.

\bibitem{Yagishita2005}
Akira Yagishita, Kouichi Hosaka, and Jun-Ichi Adachi.
\newblock Photoelectron angular distributions from fixed-in-space molecules.
\newblock {\em Journal of Electron Spectroscopy and Related Phenomena},
  142(3):295 -- 312, 2005.

\bibitem{Hockett2014}
P.~Hockett, M.~Wollenhaupt, C.~Lux, and T.~Baumert.
\newblock {Complete Photoionization Experiments via Ultrafast Coherent Control
  with Polarization Multiplexing}.
\newblock {\em Physical Review Letters}, 112(22):223001, June 2014.

\bibitem{Hockett2015b}
Paul Hockett, Christian Lux, Matthias Wollenhaupt, and Thomas Baumert.
\newblock {Maximum information photoelectron metrology}.
\newblock {\em In preparation}, 2015.

\bibitem{dielsandrudolph}
Jean-Claude Diels and Wolfgang Rudolph.
\newblock {\em Ultrashort Laser Pulse Phenomena}.
\newblock Academic Press, Oxford, UK, second edition edition, 2006.

\bibitem{Wollenhaupt2009a}
M.~Wollenhaupt, M.~Krug, J.~K\"{o}hler, T.~Bayer, C.~Sarpe-Tudoran, and
  T.~Baumert.
\newblock {Photoelectron angular distributions from strong-field coherent
  electronic excitation}.
\newblock {\em Applied Physics B}, 95(2):245--259, February 2009.

\bibitem{Note1}
For completeness we note that in the presence of resonances at the 1-photon
  level, the bound-bound transitions would look identical within a single
  active electron model, apart from taking on specific, well-defined values of
  $n$. In the case where several resonant states, e.g. high-lying Rydbergs,
  were within the laser bandwidth the dependence of the magnitudes and phases
  on $n$ would be significant. Pertinent examples of this type of effect in a
  multi-photon ionization scheme can be found in refs. \cite
  {Krug2009,Wilkinson2014}.

\bibitem{Park1996}
Hongkun Park and Richard~N. Zare.
\newblock {Molecular-orbital decomposition of the ionization continuum for a
  diatomic molecule by angle- and energy-resolved photoelectron spectroscopy.
  I. Formalism}.
\newblock {\em The Journal of Chemical Physics}, 104(12):4554, 1996.

\bibitem{Yang1948}
C.~Yang.
\newblock {On the Angular Distribution in Nuclear Reactions and Coincidence
  Measurements}.
\newblock {\em Physical Review}, 74(7):764--772, October 1948.

\bibitem{Continetti2001}
R~E Continetti.
\newblock {Coincidence spectroscopy.}
\newblock {\em Annual review of physical chemistry}, 52:165--92, January 2001.

\bibitem{Reid2012}
Katharine~L. Reid.
\newblock Photoelectron angular distributions: developments in applications to
  isolated molecular systems.
\newblock {\em Molecular Physics}, 110(3):131--147, February 2012.

\bibitem{Hockett2013}
Paul Hockett, Enrico Ripani, Andrew Rytwinski, and Albert Stolow.
\newblock {Probing ultrafast dynamics with time-resolved multi-dimensional
  coincidence imaging: butadiene}.
\newblock {\em Journal of Modern Optics}, 60(17):1409--1425, October 2013.

\bibitem{Wollenhaupt2009}
M.~Wollenhaupt, M.~Krug, J.~K\"{o}hler, T.~Bayer, C.~Sarpe-Tudoran, and
  T.~Baumert.
\newblock {Three-dimensional tomographic reconstruction of ultrashort free
  electron wave packets}.
\newblock {\em Applied Physics B}, 95(4):647--651, April 2009.

\bibitem{Smeenk2009}
C~Smeenk, L~Arissian, a~Staudte, D~M Villeneuve, and P~B Corkum.
\newblock {Momentum space tomographic imaging of photoelectrons}.
\newblock {\em Journal of Physics B: Atomic, Molecular and Optical Physics},
  42(18):185402, September 2009.

\bibitem{Hockett2010}
Paul Hockett, Michael Staniforth, and Katharine~L. Reid.
\newblock {Photoelectron angular distributions from rotationally state-selected
  NH 3 (B 1 E"): dependence on ion rotational state and polarization geometry}.
\newblock {\em Molecular Physics}, 108(7-9):1045--1054, April 2010.

\bibitem{Wollenhaupt2002}
M.~Wollenhaupt, a.~Assion, D.~Liese, Ch. Sarpe-Tudoran, T.~Baumert, S.~Zamith,
  M.~Bouchene, B.~Girard, a.~Flettner, U.~Weichmann, and G.~Gerber.
\newblock {Interferences of Ultrashort Free Electron Wave Packets}.
\newblock {\em Physical Review Letters}, 89(17):1--4, October 2002.

\bibitem{Wollenhaupt2013}
Matthias Wollenhaupt, Christian Lux, Marc Krug, and Thomas Baumert.
\newblock {Tomographic reconstruction of designer free-electron wave packets.}
\newblock {\em Chemphyschem : a European journal of chemical physics and
  physical chemistry}, 14(7):1341--9, May 2013.

\bibitem{hockettThesis}
Paul Hockett.
\newblock {\em Photoionization Dynamics of Polyatomic Molecules}.
\newblock PhD thesis, University of Nottingham, 2009.

\bibitem{Note2}
In this case parameters to allow for rotation of the image in the detector
  plane were not included, but in general could also be included.

\bibitem{bevington}
Philip~R. Bevington and D.~Keith Robinson.
\newblock {\em Data Reduction and Error Analysis for the Physical Sciences}.
\newblock McGraw-Hill, New York, 2nd edition, 1992.

\bibitem{Schmidtke2000}
B~Schmidtke, M~Drescher, N~a Cherepkov, and U~Heinzmann.
\newblock {On the impossibility to perform a complete valence-shell
  photoionization experiment with closed-shell atoms}.
\newblock {\em Journal of Physics B: Atomic, Molecular and Optical Physics},
  33(13):2451--2465, July 2000.

\bibitem{Wollenhaupt2005b}
M~Wollenhaupt, A~Pr\"{a}kelt, C~Sarpe-Tudoran, D~Liese, and T~Baumert.
\newblock {Strong field quantum control by selective population of dressed
  states}.
\newblock {\em Journal of Optics B: Quantum and Semiclassical Optics},
  7(10):S270--S276, October 2005.

\bibitem{Wollenhaupt2005}
M~Wollenhaupt, V~Engel, and T~Baumert.
\newblock {Femtosecond laser photoelectron spectroscopy on atoms and small
  molecules: prototype studies in quantum control.}
\newblock {\em Annual review of physical chemistry}, 56:25--56, January 2005.

\bibitem{Wollenhaupt2006}
M.~Wollenhaupt, A.~Pr\"{a}kelt, C.~Sarpe-Tudoran, D.~Liese, T.~Bayer, and
  T.~Baumert.
\newblock {Femtosecond strong-field quantum control with sinusoidally
  phase-modulated pulses}.
\newblock {\em Physical Review A - Atomic, Molecular, and Optical Physics},
  73(6):063409, June 2006.

\bibitem{Yin1992}
Yi-Yian Yin, Ce~Chen, D.~S. Elliott, and A.~V. Smith.
\newblock Asymmetric photoelectron angular distributions from interfering
  photoionization processes.
\newblock {\em Physical Review Letters}, 69(16):2353--2356, Oct 1992.

\bibitem{Yin1995}
Yi-Yian Yin, D.~S. Elliott, R.~Shehadeh, and E.~R. Grant.
\newblock Two-pathway coherent control of photoelectron angular distributions
  in molecular no.
\newblock {\em Chemical Physics Letters}, 241(5-6):591 -- 596, 1995.

\bibitem{Wang2001}
Zheng-Min Wang and D.~S. Elliott.
\newblock Determination of the phase difference between even and odd continuum
  wave functions in atoms through quantum interference measurements.
\newblock {\em Physical Review Letters}, 87(17):173001, Oct 2001.

\bibitem{Fiss2000}
Ja~Fiss, a~Khachatrian, K~Truhins, L~Zhu, Rj~Gordon, and T~Seideman.
\newblock {Direct observation of a breit-wigner phase of a wave function}.
\newblock {\em Physical review letters}, 85(10):2096--9, September 2000.

\bibitem{Lucchese2010}
Robert~R. Lucchese, Raffaele Montuoro, Konstantinos Kotsis, Motomichi Tashiro,
  Masahiro Ehara, John~D. Bozek, Aloke Das, April Landry, Jeff Rathbone, and
  E.D. Poliakoff.
\newblock {The effect of vibrational motion on the dynamics of shape resonant
  photoionization of BF 3 leading to the state of}.
\newblock {\em Molecular Physics}, 108(7-9):1055--1067, April 2010.

\bibitem{Lopez-Dominguez2012}
J.a. L\'{o}pez-Dom\'{\i}nguez, David Hardy, Aloke Das, E.D. Poliakoff, Alex
  Aguilar, and Robert~R. Lucchese.
\newblock {Mechanisms of Franck-Condon breakdown over a broad energy range in
  the valence photoionization of N2 and CO}.
\newblock {\em Journal of Electron Spectroscopy and Related Phenomena},
  185(8-9):211--218, September 2012.

\bibitem{Hockett2007}
Paul Hockett and Katharine~L Reid.
\newblock {Complete determination of the photoionization dynamics of a
  polyatomic molecule. II. Determination of radial dipole matrix elements and
  phases from experimental photoelectron angular distributions from A1Au
  acetylene.}
\newblock {\em The Journal of chemical physics}, 127(15):154308, October 2007.

\bibitem{Hockett2010a}
Paul Hockett, Michael Staniforth, and Katharine~L Reid.
\newblock {Photoionization dynamics of ammonia (B(1)E''): dependence on
  ionizing photon energy and initial vibrational level.}
\newblock {\em The journal of physical chemistry. A}, 114(42):11330--6, October
  2010.

\bibitem{Wang2014}
Kwanghsi Wang, Vincent McKoy, Paul Hockett, and Michael~S. Schuurman.
\newblock {Time-Resolved Photoelectron Spectra of CS2: Dynamics at Conical
  Intersections}.
\newblock {\em Physical Review Letters}, 112(11):113007, March 2014.

\bibitem{Hockett2011}
Paul Hockett, Christer~Z. Bisgaard, Owen~J. Clarkin, and Albert Stolow.
\newblock {Time-resolved imaging of purely valence-electron dynamics during a
  chemical reaction}.
\newblock {\em Nature Physics}, 7(8):612--615, April 2011.

\bibitem{Wigner1955}
Eugene Wigner.
\newblock {Lower Limit for the Energy Derivative of the Scattering Phase
  Shift}.
\newblock {\em Physical Review}, 98(1):145--147, April 1955.

\bibitem{DeCarvalho2002}
C.A.A. de~Carvalho and H.M. Nussenzveig.
\newblock {Time delay}.
\newblock {\em Physics Reports}, 364(2):83--174, June 2002.

\bibitem{Klunder2011}
K.~Kl\"{u}nder, J.~M. Dahlstr\"{o}m, M.~Gisselbrecht, T.~Fordell, M.~Swoboda,
  D.~Gu\'{e}not, P.~Johnsson, J.~Caillat, J.~Mauritsson, A.~Maquet,
  R.~Ta\"{\i}eb, and A.~L'Huillier.
\newblock {Probing Single-Photon Ionization on the Attosecond Time Scale}.
\newblock {\em Physical Review Letters}, 106(14):1--4, April 2011.

\bibitem{Dahlstrom2012}
J~M Dahlstr\"{o}m, A~L'Huillier, and A~Maquet.
\newblock {Introduction to attosecond delays in photoionization}.
\newblock {\em Journal of Physics B: Atomic, Molecular and Optical Physics},
  45(18):183001, September 2012.

\bibitem{Muller2002}
H.~G. Muller.
\newblock {Reconstruction of attosecond harmonic beating by interference of
  two-photon transitions}.
\newblock {\em Applied Physics B: Lasers and Optics}, 74:17--21, 2002.

\bibitem{Hockett2015c}
Paul Hockett, Eugene Frumker, David~M Villeneuve, and Paul~B Corkum.
\newblock {Time delay in molecular photoionization}.
\newblock {\em In preparation}, 2015.

\bibitem{Lux2015}
Christian Lux, Matthias Wollenhaupt, Cristian Sarpe, and Thomas Baumert.
\newblock {Photoelectron Circular Dichroism of Bicyclic Ketones from
  Multiphoton Ionization with Femtosecond Laser Pulses}.
\newblock {\em ChemPhysChem}, 16(1):115--137, January 2015.

\bibitem{Krug2009}
M~Krug, T~Bayer, M~Wollenhaupt, C~Sarpe-Tudoran, T~Baumert, S~S Ivanov, and N~V
  Vitanov.
\newblock {Coherent strong-field control of multiple states by a single chirped
  femtosecond laser pulse}.
\newblock {\em New Journal of Physics}, 11(10):105051, October 2009.

\bibitem{Wilkinson2014}
Iain Wilkinson, Andrey~E Boguslavskiy, Jochen Mikosch, Julien~B Bertrand,
  Hans~Jakob W\"{o}rner, David~M Villeneuve, Michael Spanner, Serguei
  Patchkovskii, and Albert Stolow.
\newblock {Excited state dynamics in SO2. I. Bound state relaxation studied by
  time-resolved photoelectron-photoion coincidence spectroscopy.}
\newblock {\em The Journal of chemical physics}, 140(20):204301, May 2014.

\end{thebibliography}

\end{document}